\documentclass[10pt, aps, prc, showkeys, longbibliography, nofootinbib]{revtex4-1}
\usepackage{epsfig,graphicx,float,color}
\usepackage{amssymb}
\usepackage{amsmath}
\usepackage{longtable}
\usepackage{setspace}
\usepackage{comment}
\usepackage[ulem=normalem,defaultcolor=red,commandnameprefix=always]{changes}
\usepackage{statmath}
\usepackage[braket, qm]{qcircuit} 

\usepackage{hyperref}
\hypersetup{colorlinks=true,linkcolor=blue,citecolor=blue}

\begin{document}
\title{The extended Lipkin model: proposal for implementation in a quantum platform and machine learning analysis of its phase diagram}
\author{S.~Baid$^{1}$, A.\ S\'aiz$^{1,2}$, L.\ Lamata$^{1}$, P.~P\'erez-Fern\'andez$^{2}$, A.M.~Romero$^{3,4,a}$
, A.~Rios$^{3,4}$, J.M.~Arias$^{1}$ and J.E.~Garc\'{\i}a-Ramos$^{5,6}$}
\affiliation{
  $^1$Departamento de F\'isica At\'omica, Molecular y Nuclear, Facultad de F\'isica\text{,} Universidad de Sevilla, Apartado 1065, E-41080 Sevilla, Spain\\
  $^2$Departamento de Física Aplicada III\text{,} Escuela Técnica Superior de Ingeniería, Universidad de Sevilla, E-41092 Sevilla, Spain.\\  
  $^3$ Departament de Física Quàntica i Astrofísica (FQA), Universitat de Barcelona (UB), c.\ Martí i Franqués, 1, 08028 Barcelona, Spain\\
  $^4$ Institut de Ciències del Cosmos (ICCUB), Universitat de Barcelona (UB), c.\ Martí i Franqués, 1, 08028 Barcelona, Spain\\
  $^5$Departamento de  Ciencias Integradas y Centro de Estudios Avanzados en F\'isica, Matem\'atica y Computaci\'on, Universidad de Huelva, 21071 Huelva, Spain.\\
  $^6$Instituto Carlos I de F\'{\i}sica Te\'orica y Computacional,  Universidad de Granada, Fuentenueva s/n, 18071 Granada, Spain\\
  $^a$New address: Fujitsu Research of Europe, antonio.marquezromero@fujitsu.com  
}
\begin{abstract}
\begin{description}
\item [Background]
In recent years, the implementation of Nuclear Physics models in quantum computers has emerged as a promising and novel area of research. Simultaneously, the study of quantum shape phase transitions in nuclear models has gained significant attention. Specifically, the phase diagram of the Interacting Boson Approximation (IBA) has been extensively explored, particularly in connection with 
large-particle-number-limit considerations. Interestingly, the Extended Lipkin Model (ELM) serves as a valuable alternative for mimicking the IBA phase diagram and holds the advantage of being more straightforward to implement within a quantum computing platform.

\item [Purpose]
We explore the ELM and provide: i) calculations of the ground state energy using a variational quantum eigensolver; ii) a comprehensive formulation for implementing the dynamics of the ELM within a quantum computing platform, enabling the experimental exploration of the IBA phase diagram for systems with a small number of particles; and iii) a determination of the phase diagram of the model using different Machine Learning (ML) methods. We note that in the ELM, unlike the usual Lipkin model, both first- and second-order quantum shape phase transitions take place depending on the model parameters.

\item [Method]
 Initially, we employ the Adaptive Derivative-Assembled Pseudo-Trotter ansatz Variational Quantum Eigensolver (ADAPT-VQE) to calculate the ground-state energy of the ELM.  Next, we introduce the essential formulation and procedures required to implement this model effectively in a quantum computing environment. Finally, we use ML techniques to identify the different phases and critical points of the ELM. 

\item [Results]
We successfully reproduce the ground-state energy of the ELM across the complete phase space of the model using the ADAPT-VQE algorithm. We provide the necessary framework for implementing the ELM in a quantum computing platform, ensuring that the model can be executed with controlled errors. Finally, we obtain meaningful ML predictions that allow us to determine the phase diagram of the model. 

\item [Conclusions]
Our findings offer compelling evidence that the implementation of a nuclear model like the ELM in a quantum computing environment is not only feasible but can also be achieved with manageable error rates. This realization opens the door to detailed experimental investigations of the phase diagram of the ELM (and indirectly of the IBA) in a quantum computer, further advancing our understanding of quantum shape phase transitions and nuclear structure.
\end{description}
\end{abstract}

\keywords{Quantum Platforms \quad Nuclear Models \quad ADAPT-VQE \quad Quantum Shape Phase Transitions \quad Interacting Boson Approximation \quad Extended Lipkin Model \quad Machine Learning}

\date{\today}

\maketitle

\section{Introduction}
\label{sec-intro}
A phase is a specific state of matter characterized by a well-defined structure and physical properties. Phase transitions correspond to abrupt changes in some physical properties of a system as it passes from one phase to the other \cite{Land69,Stan1987,NyO2010}. Phase transitions are associated with sudden shifts in the internal energy of the system that occur at specific values of a control parameter, that is, a physical magnitude that can be changed externally to the system. The point of change is called critical point and, in addition to changes in energy, qualitative changes in other physical quantities, called order parameters, also appear. Order parameters behave differently depending on the phase, so they can potentially be used to distinguish among different phases. Phase transitions are well defined and characterized in macroscopic systems~\cite{Land69,Stan1987,NyO2010}.

When one looks into mesoscopic quantum systems, a natural question is whether one can find similar phase transitions to the macroscopic ones or not. One may also wonder whether the same concepts introduced to study classical phase transitions apply at the quantum level. 
The study of Quantum Phase Transitions (QPTs) is a popular and widely discussed topic that encompasses various branches of quantum many-body physics \cite{NyO2010,Sach2011,Vojta2003,Carr2010}. In recent years, there has been a growing interest in investigating low-dimensional lattice models due to advances in quantum computing \cite{Kokail2019,Bauer2023,Faus2024}. In addition, concepts from quantum information theory have been used to characterize quantum critical phenomena \cite{vidal2003,song2017,liu2017}. On a different note, there has been a resurgence in the examination of structural changes in finite-size systems, where early signs of transitions can be observed \cite{Iach04a}.

\begin{figure}[t]
  \centering
    \centering
  \includegraphics[width=0.60\textwidth]{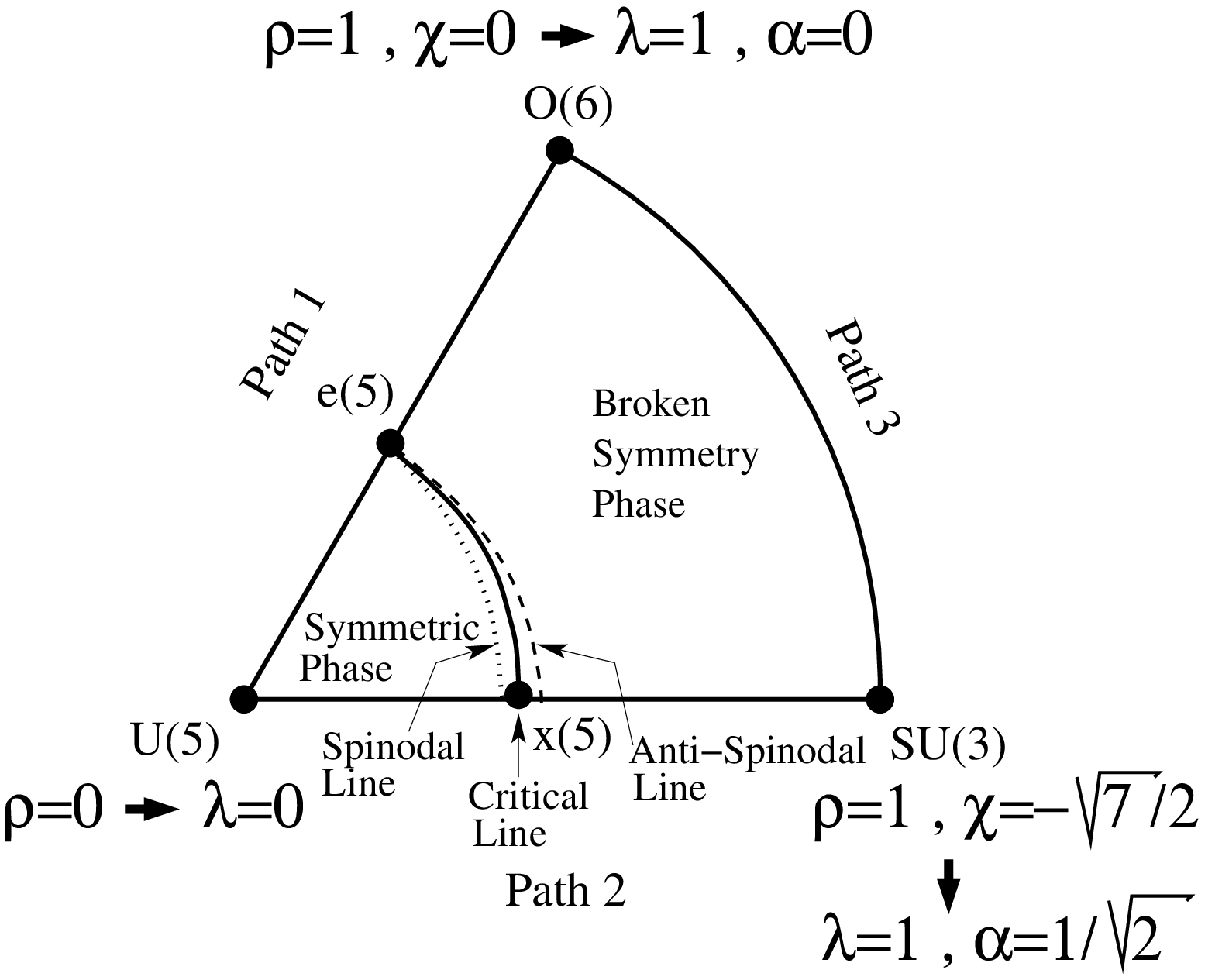}
  \caption{Phase diagram (or Casten triangle) for the IBA Hamiltonian (\ref{HCQF2}) with control parameters $\rho$ and $\chi$. The dots at the vertices mark the three dynamical symmetries in the model (U(5), O(6) and SU(3)) and the dots in between the U(5)-O(6) and the U(5)-SU(3) lines mark the two critical point symmetries (e(5) and x(5)) in the model. In addition, the relation with the ELM phase diagram defined in terms of the control parameters $\lambda$ and $\alpha$ (\ref{eq:hamiltonian}) is shown.}
  \label{phaseDIBM}
\end{figure}

In the realm of atomic nuclei, the Interacting Boson Approximation (IBA) \cite{iach87} provides a straightforward, yet detailed, framework to study first- and second-order phase transitions. In the IBA, there are three limiting situations, called dynamical symmetries, in which exact analytical solutions to the complex nuclear many-body problem exist. These are called U(5), SU(3) and O(6) dynamical symmetries. Remarkably, each of these symmetries corresponds to a specific shape in terms of the usual nuclear deformation parameters: $\beta$, that measures the deviation from sphericity, and $\gamma$, that measures the deviation from axial symmetry. U(5) corresponds to spherical shapes, SU(3) to axial prolate deformed shapes and O(6) to $\gamma$-unstable deformed shapes \cite{iach87}. Thus, the phase space of the IBA is usually represented as a triangle such that each dynamical symmetry is located at one vertex \cite{Caste1981}. We show this so-called Casten triangle in Fig.~\ref{phaseDIBM}. In this figure, the critical line delineating the phase transition from spherical to deformed shapes is depicted by a solid arc connecting the e(5) critical point (representing a second-order phase transition on the U(5)-O(6) line) and the x(5) critical point (representing a first-order phase transition on the U(5)-SU(3) line). The dashed line (anti-spinodal) and dotted line (spinodal) demarcate the regions where shape coexistence occurs within the model.

At the mean-field limit, there are two phases in the model: a spherical phase and a deformed one. The transitions between these phases are of first or second order depending on the path followed in the model parameter space. For instance, the transition from the spherical U(5) dynamical symmetry to the deformed $\gamma$-unstable O(6) dynamical symmetry is exactly solvable. This corresponds to the boson pairing model initially solved by Richardson~\cite{Rich1968}, and later applied to the IBA \cite{Duke2001a,Duke2001b}. These methods provide access to solutions for a very large number of bosons along this line.  Graphically, this line corresponds to one side of the Casten triangle $\mbox{U(5)} \rightarrow \mbox{O(6)}$. This is the only integrable line in the IBA phase diagram and there, a second order QPT exists. 
The interior region of the Casten triangle, as well as the other two sides, remain beyond the scope of large-scale numerical studies that are limited by the capabilities of standard IBA codes, which can handle systems with approximately $100$ bosons \cite{Bare2010}. In all this area a critical line of first order QPTs appears. 
Other phase diagrams corresponding to algebraic models  have been studied, e.g., the Agassi model~\cite{Garc2018}. The model has been implemented in a quantum platform and the phases of the system was determined resorting to the analysis of the time evolution of a properly chosen operator~\cite{Perez_Fernandez_2022}. In a further analysis, the phase diagram has been determined using Machine Learning (ML) methods \cite{Saiz_2022}.

Strictly speaking, QPTs only occur in macroscopic systems. One may consider different avenues to connect QPTs to the physics of few-body systems. First, we may want to gain access to numerical simulations that can address low as well as large particle numbers (large-$N$ limit). A second complementary approach would be to encode the model in a quantum platform, where large-$N$ scalings may be tamed down in the long term. These two research avenues are particularly challenging for the IBA, but other models may not suffer from the same limitations. The purpose of this paper is precisely to overcome the limitations of the two analyses suggested above with a model that is similar, but not as challenging, as the IBA. 


We use here the extended Lipkin model (ELM) proposed in Ref.~\cite{Vida2006,Roma2021}. In the large-$N$ limit, the ELM mimics the IBA-shape phase diagram. Moreover, the ELM is simpler than the IBA, and it can be formulated and solved in terms of spin-$\frac{1}{2}$ Pauli operators. Consequently, the ELM can naturally be implemented in a quantum platform with standard qubit registers. Furthermore, the numerical solution to the ELM is 
accessible classically for a wide range of particle numbers. In this sense, the ELM is a perfect test-bed for QPTs employing both classical and quantum techniques. In this paper, we exploit a variety of ML and quantum computing methods to assess the different phases of the model at finite $N$ and to demonstrate the usefulness of the model for a wide range of applications. We ultimately hope to prove that this model has the potential to become a new standard  for QPT studies in quantum devices. 


The paper is organized as follows. In Sect.\ \ref{Sect-2}, we provide a short revision of the usual Lipkin model, the IBA and the ELM. We specifically discuss the relation between the IBA and the ELM phase diagrams. In Sect.\ \ref{Sect-3}, we present an ADAPT-VQE implementation for the ELM, providing benchmarks with ground-state energies for a wide range of values of the control parameters.  
We introduce a proposal for the implementation of the ELM in a quantum platform, together with an evaluation of the possible errors,  in Sect.\ \ref{Sect-4}. 
ML techniques are applied to investigate the phase diagram of the ELM in Sect.\ \ref{Sect-5}.  Finally, Sect.\ \ref{Sect-6} provides the summary and conclusions.  

\section{The models}
\label{Sect-2}

\subsection{The Lipkin model}

The Lipkin-Meshkov-Glick model (LMG) \cite{Lipk1965} (Lipkin model in short) describes a set of $N$ particles interacting with a long-range interaction. Each particle is allowed to be in a two-level system, upper and lower, separated by an energy gap. If each particle is considered to have spin $1/2$, the complete set can be represented by a collective spin $\vec S$. Defined as $\vec S = \sum_{i=1}^{N} \vec s_{i}$ where $\vec s_i$ is the spin of each individual particle. The total collective spin $S=N/2$ is the relevant sector for this work and it includes the ground state of the system.

The original Lipkin Hamiltonian \cite{Lipk1965} can be written\footnote{The original Lipkin Hamiltonian is really $H_L = \varepsilon J_z - \frac{V}{2} (J_+^2 + J_-^2) - \frac{W}{2} (J_+J_- +J_-J_+)$.
Taking $W=V$, up to a global scale factor, it can be written as,
$H_L = (1-\lambda) J_z - \frac{4 \lambda}{N} ~J_x^2$ that is, apart from a constant factor, Eq. (\ref{eq:Lipkinhamiltonian}). The factor $1/N$ is included to scale properly the relative strengths of the one- and two-body interactions.} as
\begin{equation}
  \label{eq:Lipkinhamiltonian}
  H_L= (1-\lambda) (S+S_z) - \frac{\lambda}{N} (S_+ + S_-)^2 = (1-\lambda) (S+S_z) - 4 \frac{\lambda}{N} S_x^2 ~,
\end{equation}
where $S_{\nu} = \sum_{i=1}^{N} s_{i,\nu}$ for $\nu = x,y,z$ and $S_\pm = S_x \pm i S_y$. Here, $\lambda$ is a control parameter, and the Hamiltonian can be diagonalized in the collective basis $|S M_S\rangle$, where $S=N/2$ is fixed and $M_S$ runs from $-N/2$ to $+N/2$ in unit steps and the dimension of the Hilbert space is $\mbox{dim}=N+1$. 
In this way, the complete energy spectrum can be obtained as a function of $\lambda$. 
Although the Lipkin model has a very simple structure, it displays interesting symmetry properties and phase transitions. 
Its phase diagram has been extensively studied~\cite{agassi1966validity,wahlen2017merging}. It is known that, in the form given by Eq.\ (\ref{eq:Lipkinhamiltonian}), the LMG presents two phases in the large-$N$ limit. These symmetric and non-symmetric phases are separated by a second-order shape phase transition at the critical point, $\lambda_c=1/5$.
We note that the Lipkin model has become a standard benchmark and test-bed for implementation of quantum algorithms focused on nuclear structure \cite{denis2022,denis2024,robinsavage2023,fabaI,fabaII,fabaIII}. 

\subsection{The interacting boson approximation}
\label{sec-IBA}
The interacting boson approximation (IBA) \cite{iach87} is a successful and widely employed nuclear structure model valid to describe low-lying levels of medium and heavy nuclei. The model is formulated in a second quantization formalism that includes a scalar boson ($s$ boson, $L=0$) and five quadrupole bosons ($d_\mu$ boson, $L=2$). The IBA has $u(6)$ as dynamical algebra and three specific situations, called dynamical symmetries, can be distinguished for which analytical solutions to the complex nuclear many-body problem are obtained. These dynamical symmetries are named: U(5), SU(3) and O(6). A simplified form of the IBA Hamiltonian is the so-called Consistent-Q Hamiltonian (CQH) \cite{WC1983,CW1988},
\begin{equation}
  H_{CQH} = \epsilon ~ n_d - \kappa ~ Q^{(\chi)} \cdot Q^{(\chi)} , 
  \label{HCQF}
\end{equation}
where $n_d$ is the $d-$boson number operator, 
$Q^{(\chi)} = s^\dag \tilde d + d^\dag s + \chi (d^\dag \tilde d)^{(2)}$ is the quadrupole operator and $\epsilon, \kappa$ and $\chi$ are the model parameters. The phase space of the IBA is usually represented  in the Casten triangle (Fig.~\ref{phaseDIBM}). 
Although the IBA is formulated within the context of the second quantization formalism, it is possible to develop a geometric representation of the model by utilizing the concept of coherent states and the intrinsic state formalism \cite{Gino1980a,Gino1980b,Diep1980}. This alternative approach unveils geometric shapes reminiscent of the Bohr model \cite{Bohr1975}, which are associated with the three dynamical symmetries of the IBA. Consequently, within this framework, the U(5) limit corresponds to spherical shapes, while the SU(3) 
limit leads to prolate 
axially symmetric shapes. Additionally, the O(6) limit gives rise to deformed $\gamma-$unstable shapes. 
Since the model presents several phases, studies of these and of the corresponding quantum phase transitions are of great interest.
Sometimes, to study phase transitions, the CQH in Eq.~(\ref{HCQF}) is rewritten, taking out a global scale factor, as
\begin{equation}
  H_{CQH} = (1-\rho) ~ n_d - \frac{\rho}{N}~ Q^{(\chi)} \cdot Q^{(\chi)}, 
  \label{HCQF2}
\end{equation}
where $N$ is the total boson number of the system. The specific values of $\rho$ and $\chi$ that produce the three IBA dynamical symmetries are given in Fig.~\ref{phaseDIBM}. Excellent reviews presenting the IBA in relation with its phase diagram and the corresponding shape phase transitions are Refs.\ \cite{Caste2006,Caste2007,Caste2009,Cejnar2009,Cejnar2010}.

However, it is important to recognize that QPTs manifest themselves genuinely in macroscopic systems. Consequently, to study them without using mean-field methods, one needs to consider a significantly large number of bosons, which is not always computationally feasible. 
A possible solution is to look for a simpler model with a similar phase diagram, as the one presented in the next subsection.

\begin{figure}[t!]
\centering
\begin{tabular}{cc}
a) & b) \\
\begin{minipage}{0.4\textwidth} 
  \centering
  \includegraphics[width=0.95\textwidth]{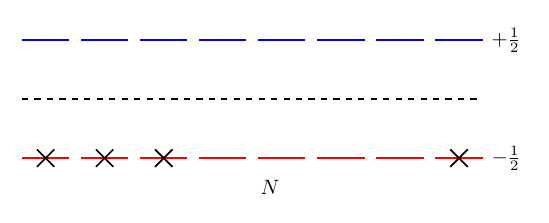}
\end{minipage}
&
\begin{minipage}{0.4\textwidth}
  \centering
  \includegraphics[width=0.95\textwidth]{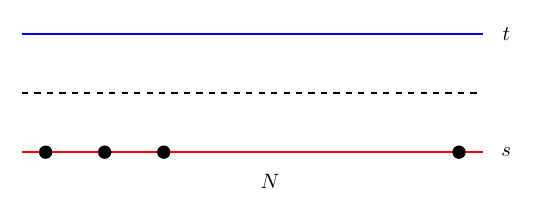}
\end{minipage}
\end{tabular}
\caption{Lipkin model in terms of fermion operators (a) and, equivalently, in terms of scalar boson operators (b). }
\label{Fig-st-Lipkin}
\end{figure}

\subsection{The extended Lipkin model}
\label{sec-elm}
The Extended Lipkin Model (ELM) \cite{Vida2006} is a simple two-level model designed to mimic the IBA phase diagram. Although the Lipkin model was originally formulated in terms of fermions or spin operators, as discussed above, it can be alternatively formulated in terms of scalar bosons using the Schwinger representation \cite{chaturvedi2006schwinger} (see Fig.~\ref{Fig-st-Lipkin}). Using this boson formulation, the ELM was introduced as a two-level  model constructed on two types of $L=0$ bosons, $s$ and $t$, with Hamiltonian
\begin{equation}
  \label{eq:hamiltonian}
  H= (1-\lambda)  ~n_t-\frac{\lambda}{N} ~ Q^{(\alpha)} \cdot Q^{(\alpha)},
\end{equation}
where the operators $n_t$, which  corresponds to the number of $t-$bosons, and $Q^{\alpha}$ are defined as
\begin{equation}
  \label{eq:Qdef}
 n_t= t^\dag t, \hspace{0.5cm} Q^{(\alpha)}=(s^\dag t + t^\dag s) + \alpha  (t^\dag t)~.
\end{equation}
$\lambda$ and $\alpha$ are two independent control parameters that play the role of $\rho$ and $\chi$ in the IBA Hamiltonian, Eq.\ (\ref{HCQF2}). The total number of bosons $N=n_s + n_t$ is a conserved quantity.

We have deliberately written the ELM two-level boson Hamiltonian in the form of Eq.~(\ref{eq:hamiltonian}) so that it resembles the QCH of Eq.~(\ref{HCQF2}). This way, it becomes clear that $s$ and $t$ play the role of the $s$ and $d$ bosons of the IBA respectively. Clearly, the difference resides in the quadrupole character of the $d$ boson that leads to the u(6) dynamical algebra of the IBA, while the ELM is described by a u(2) dynamical one. Despite this important difference, the ELM captures the main features of the phase diagram of the IBA \cite{Vida2006}. 

Using mean-field techniques similar to those used in the IBA \cite{Caste2006,Caste2007,Caste2009,Cejnar2009,Cejnar2010}, one can obtain the energy per particle surface for the ELM of Eq.~(\ref{eq:hamiltonian}) as a function of the control parameters, $\lambda$ and $\alpha$, and a variational parameter, $\beta$,
\begin{equation}
  \label{eq:ener-surf}
  \frac{E(\lambda,\alpha;\beta)}{N} = \frac{\beta^2}{(1+\beta^2)^2}~ \left[ \beta^2 (1-(1+\alpha^2) \lambda) - \beta 4 \alpha \lambda + (1-5 \lambda) \right].
\end{equation}
This allows us to obtain the classical limit of the Hamiltonian and to produce a phase diagram of the ELM model.

\medskip

The connection between the Hamiltonian in Eq.~(\ref{eq:hamiltonian}) and its form in terms of spin operators can be obtained by making the inverse Schwinger transformation, defining the collective spin $\vec S$ as:
\begin{equation}
S_{+}= t^\dag s, \quad S_{-}= s^\dag t, \quad S_{z}=\frac{1}{2}(t^\dag t- s^\dag s).
\end{equation}
Using this fermionic formulation, given a number of spin-$\frac{1}{2}$ particles, 
the sector containing the ground state is given by a collective spin $S=N/2$ and the model Hamiltonian is written as,
\begin{equation}
  \label{eq:ELipkinhamiltonian}
  H_{EL}= H_L - \frac{\lambda}{N} \left[\alpha^2 (S + S_z)^2 - 2 \alpha \big(S_x (S+S_z) + (S+S_z) S_x \big)  \right],
\end{equation}
in such a way that for $\alpha=0$ the usual Lipkin Hamiltonian, $H_L$ in Eq.\ (\ref{eq:Lipkinhamiltonian}), is recovered. This Hamiltonian can also be easily diagonalized in a basis given by $|S M_s\rangle$ with $S=N/2$.

\subsection{The phase diagram of the IBA and the ELM}
\label{sec-ph-diagram}
We employ Eq.~(\ref{eq:ener-surf}) to determine the phase diagram of the ELM Hamiltonian in 
Eq.~(\ref{eq:hamiltonian}) (or Eq.~(\ref{eq:ELipkinhamiltonian}) in the fermionic formulation). 
The corresponding ELM phase diagram mimics the IBA phase diagram depicted in Fig.~\ref{phaseDIBM} for the IBA Hamiltonian in Eq.~(\ref{HCQF2}). As a matter of fact, the connection between the parameters of both models is rather simple,
\begin{equation}
\rho\rightarrow \lambda, \qquad \chi\rightarrow -\sqrt{\frac{7}{2}} \alpha .
\end{equation}
In addition to the critical line displayed by the solid curve,
\begin{equation}
  \lambda_c=\frac{1}{5+\alpha^2} , 
  \label{lc}
\end{equation}
we show the spinodal (dotted) $\lambda_s$ and antispinodal (dashed) $\lambda_{as}$ lines, given by the expressions
\begin{equation}
\lambda_s =\frac{\alpha ^2-\sqrt{\alpha ^4+10 \alpha ^2+16}+6}{\alpha ^2+10}~, \qquad \lambda_{as}=1/5~.
  \label{spinodal}
\end{equation}

\begin{figure}
\centering
  \includegraphics[width=0.6\textwidth]{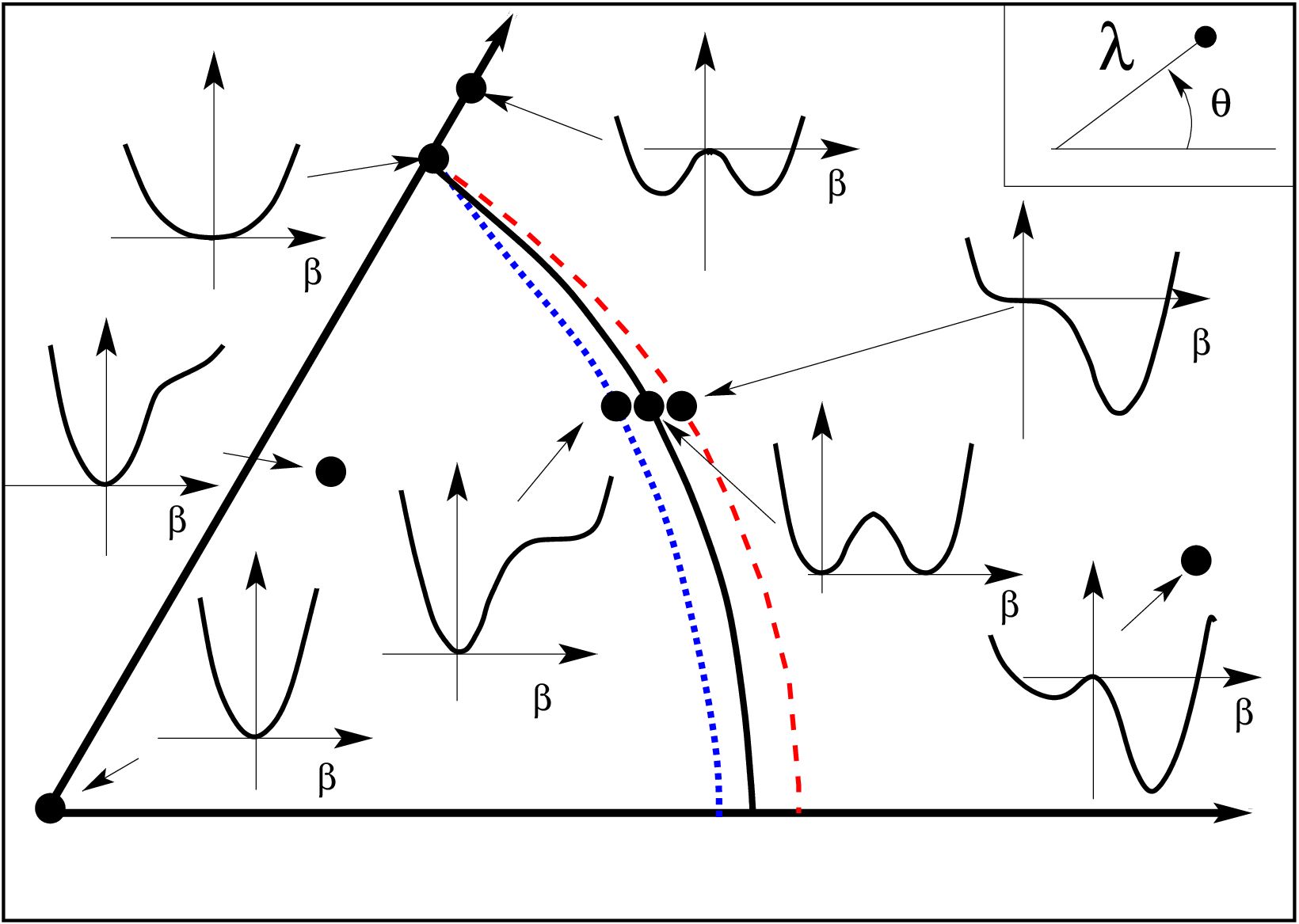}
\caption{Zoom around the critical line of the ELM phase diagram for the Hamiltonian Eq.~(\ref{eq:hamiltonian}) in the plane of the control parameters $(\lambda,\alpha)$. The value of $\alpha$ is directly connected with $\theta$ through $\alpha=\sqrt{2/3}\sin(\pi/3-\theta)$. For $\alpha=0$, the transition is second-order, whereas it is first-order otherwise. We show the critical line of Eq.\ (\ref{lc}) as a solid line, and the 
spinodal and anti-spinodal lines of Eq.\ (\ref{spinodal}) in dotted and dashed lines, respectively. 
We provide schematic representations of the energy surfaces as a function of the deformation parameter, $\beta$, in the different regions of the phase diagram.}
\label{phaseD}
\end{figure}

We provide a schematic representation of the phase diagram in Fig.~\ref{phaseD}.
The spinodal and antispinodal lines in this diagram
define the region where the energy surface presents two minima: a spherical one and a deformed one, indicating a coexistence of both phases. In the symmetric phase, there is a single energy minimum at the origin ($\beta=0$). When the spinodal line is reached, a second, non-symmetric minimum develops for  $\beta \neq 0$. At the critical line, the symmetric and the non-symmetric minima are degenerate. Then, at the anti-spinodal line the symmetric minimum at $\beta=0$ disappears to become an inflection point. From there on, the $\beta=0$  point becomes a maximum and there are only non-symmetric deformed minima. 

The energy functional given by Eq.\ (\ref{eq:ener-surf}) is a case of the cusp catastrophe whose germ is $x^4$ \cite{Gilm1981}. 
For $\alpha=0$, which corresponds to the usual Lipkin model, there is an isolated point of second-order phase transition, with a $\beta^4$ behavior at the origin. In that case, the spinodal, anti-spinodal and critical points coincide at the critical value, $\lambda_c=1/5$. For $\alpha \neq 0$, the phase transition changes its character to first order with the critical point given by Eq.~(\ref{lc}). Here, both minima (one spherical and one deformed) are degenerate and there is a coexistence region around the critical point, $\lambda_c$, between $\lambda_{as}$ and $\lambda_s$. Moreover, for $\lambda>\lambda_{as}$, two non-symmetric minima coexist. 

\subsection{ELM numerical calculations}
\label{sec-ELM-paths}
Having established the similarity of the IBA and the ELM models at the mean field level, we now proceed to present numerical ELM calculations for a finite particle number, $N$. We aim to show that precursors of the phase transitions are clearly observed even for small systems. All calculations are done for $N=6$.

\begin{figure}[t]
\centering
    \includegraphics[width=0.75\textwidth]{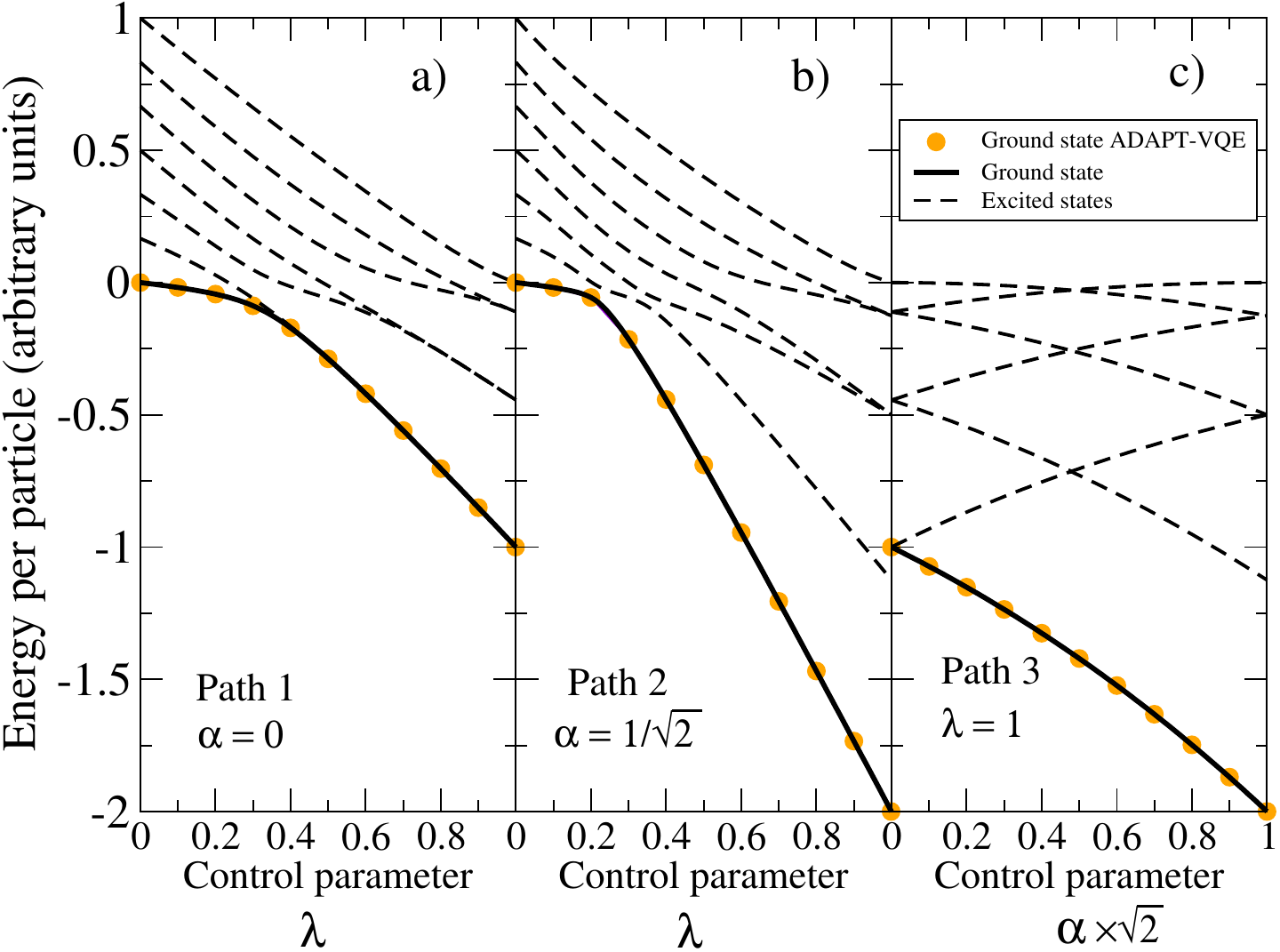}
\caption{Correlation energy diagram for the ELM for three different paths: a) path 1, $\alpha=0$ and $\lambda$ changing from 0 to 1 (second order phase transition), b) path 2, for $\alpha=1/\sqrt{2}$ and $\lambda$ changing from 0 to 1 (first order phase transition), and c) path 3, for $\lambda=1$ and $\alpha$ changing from 0 to $1/\sqrt{2}$ (no phase transition). Calculations are performed for $N=6$, and the represented magnitudes are per particle ($E/N$). Different line styles are used for different levels (full line for the ground state and broken lines for the excited states). Full (orange) dots are the results for the ground-state energy obtained with the ADAPT-VQE approximate calculations (see Section \ref{Sect-3}).}
\label{energies-Fig}
\end{figure}

We select three distinct paths that represent various regions of the phase diagram.
Path 1 follows the $\alpha=0$ line, corresponding to the standard Lipkin model, which exhibits a second-order phase transition at the mean-field level. Path 2 traverses the $\alpha=1/\sqrt{2}$ line, associated with a first-order phase transition. Finally, path 3 keeps $\lambda=1$ while varying $\alpha$ as the control parameter.
Figures \ref{energies-Fig}(a), \ref{energies-Fig}(b), and \ref{energies-Fig}(c) illustrate the correlation energies along these paths in the ELM phase diagram.

In the large-$N$ limit, the ground state energy per particle displays a change in slope at the critical point, $\lambda_c$ of Eq.~(\ref{lc}). This transition is gradual for $\alpha=0$ (path 1) due to the second-order nature of the phase transition. For $\alpha\neq 0$, the transition is more abrupt, particularly for $\alpha=1/\sqrt{2}$ (path 2), as a result of a first-order phase transition. Despite the finite size effects in our simulations for a small system with $N=6$, Fig.~\ref{energies-Fig} demonstrates precursors of a QPT in the ground state energy around $\lambda_c$ for paths 1 and 2, depicted in panels (a) and (b). Notably, the slope change near $\lambda_c$ is smooth for path 1 (panel a), indicative of a second-order phase transition, and more abrupt for path 2 (panel b), suggesting a first-order phase transition. In contrast, no indication of a QPT is observed for path 3 (panel c). It is worth noting that the apparent crossings observed for some excited states in panel (c) are not genuine, but rather avoided crossings, as the mechanism of level repulsion is at play in this scenario.



\begin{figure}[t]
\centering
\includegraphics[width=0.75\linewidth]{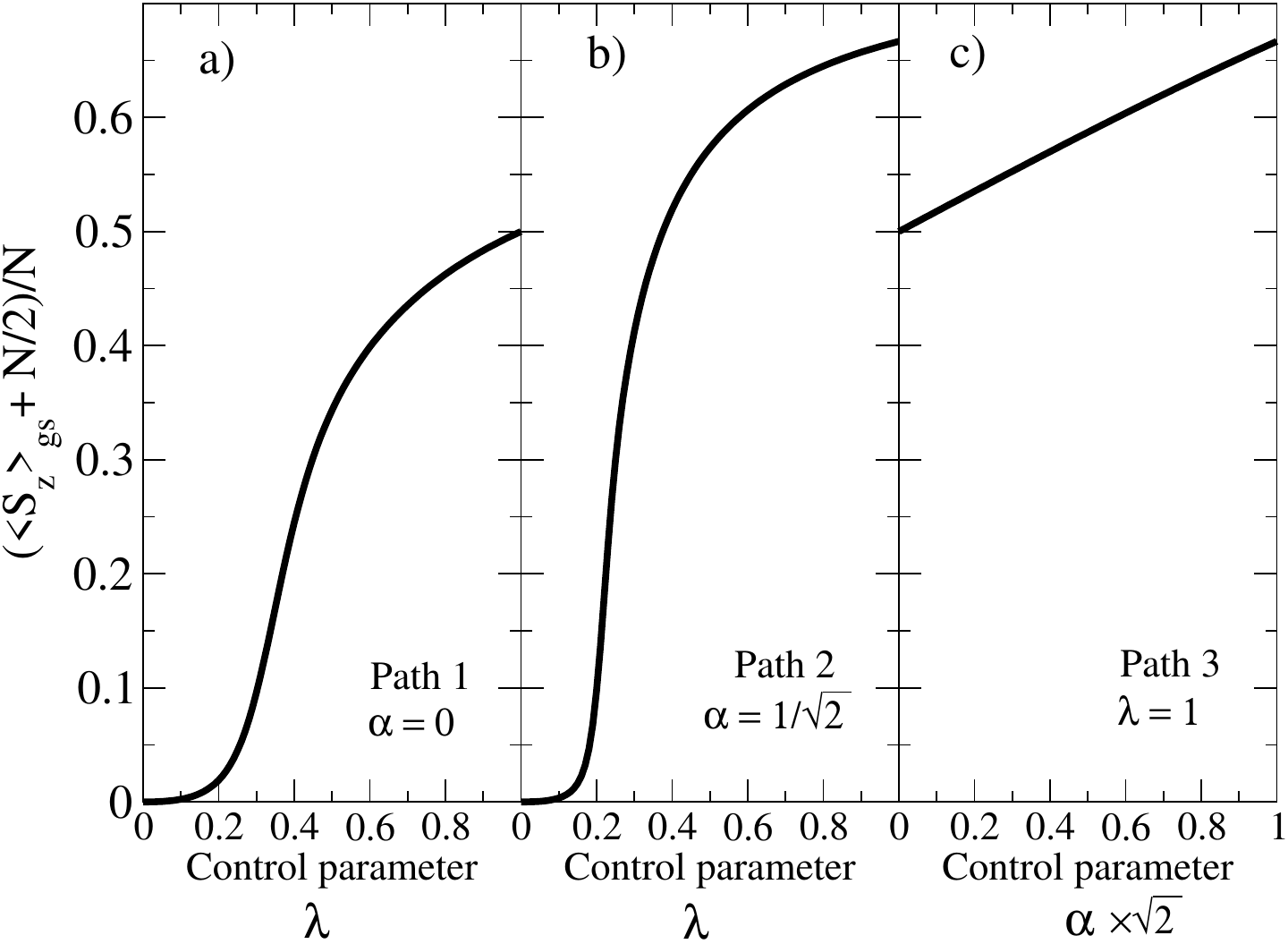}
\caption{Expectation value of $S_z+N/2$, normalized to N, for the ELM ground state along the three paths discussed in the text as a function of the control parameter. Calculations are done for $N=6$.}
\label{fig-ELM-order-param}
\end{figure}

In addition to the energy of the system, there are other physical properties that reflect the change of phase. 
For the ELM, the expectation value of $(S_z+S)$ in the ground state can be considered as an order parameter that, in the large-$N$ limit, is zero in the symmetric phase, and different from zero in the broken one. Panels a), b) and c) in Fig.~\ref{fig-ELM-order-param} represent this order parameter, normalized to $N$, for $N=6$ and the same three paths discussed in the preceding figure. 
For finite particle number, $N$, the sudden change of behaviour in this order parameter is not exactly at the mean-field critical value of the control parameter.
Also for this observable, precursors for the phase transition are clearly observed in the ground state for paths 1 and 2 in the vicinity of $\lambda_c$, Eq.~(\ref{lc}). 
In contrast, we do not see change in behavior along path 3, where the system is always in the broken phase. In addition, in Fig.~\ref{fig-ELM-order-param} the change in the order parameter at the critical point is more abrupt for path 2 than for path 1. This is due to the fact that, in the large-$N$ limit, the transition in path 1 is of second order (continuous change in the order parameter at the critical point), while in path 2, it is of first order (discontinuous change in the order parameter at the critical point). 

\section{ADAPT-VQE algorithm for the ELM}

We have so far discussed the phase diagram of the ELM. We have illustrated the richness of the phase diagram of this model, which is reflected in the properties of even relatively small systems. We now turn to discuss how simulations for such small systems may potentially be performed in quantum devices. 

\label{Sect-3}
\subsection{The ADAPT-VQE algorithm } 

The ADAPT-VQE algorithm \cite{Grim2019,Romero2022} is a method within the class of Variational Quantum Eigensolver (VQE) approaches \cite{peruzzo2014variational,Tilly2022}. VQEs aim at solving variational problems, such as the ground state of quantum many-body systems, by 
parametrizing their 
wave functions in terms of quantum circuits, and minimizing the corresponding energies employing classical algorithms. The hope is that such hybrid methods may provide a quantum advantage over classical simulations~\cite{peruzzo2014variational,Tilly2022}

ADAPT-VQE  was originally devised to calculate bond dissociation curves in molecules~\cite{Grim2019}, but has since been employed to study other physical systems like the Lipkin model~\cite{Romero2022} or nuclei~\cite{Pere2023}. 
The methods is iterative and starts from an initial reference trial state, $|\rm{ref}\rangle$. This state is updated by means of encoded operators in a previously-defined operator pool. Thus, at iteration $n$, the resulting quantum state $|n\rangle$ is given by a product
\begin{equation}
|n\rangle = e^{i\theta_n A_n}|n-1\rangle = \prod_{k=1}^{n} e^{i\theta_k A_k}|\text{ref}\rangle.
\end{equation}
Here, $\theta_k$ with $k=1, \cdots, n$ are $n$ variational parameters
associated to the $n$ operators, $A_k$. 
The optimal values of these parameters are obtained by minimization of the energy surface resulting from the expectation value of this state with the Hamiltonian. 
At iteration $n-1$, the algorithm must decide which operator to choose at iteration $n$. 
The gradient at $\theta_k=0$ can be written in a compact expression as \cite{Grim2019} 
\begin{equation}
\frac{{\partial E^{(n)}}}{{\partial \theta_k}}\bigg|_{\theta_k=0} = i \langle n | [H, A_k] | n \rangle.
\end{equation}
The algorithm selects the operator $A_{n}$ that produces the largest gradient, and thus is expected to have the optimal descent in the energy surface. Compared to approaches without adaptive gradients, this should allow for a faster approach to the real energy minimum, potentially avoiding plateaus~\cite{Grim2023}. We note that the operators $A_k$
are selected from a common pool, and may be repeated along the iterative minimization. 

Within ADAPT, quantum circuits can be used to construct the states and measure the commutators. The variational parameters are optimized at each iteration minimizing the total energy surface, on a classical computer, aided by measurements in the quantum computer \cite{Pere2023}. 
For the purpose of evaluating its performance, we do not operate explicitly with quantum circuits, but rather  simulate the evolution of the algorithm on a classical computer. 

We intend to show that the formalism developed in Ref.~\cite{Romero2022} for the standard Lipkin model is also valid for the ELM. We note this is indeed the case, even though the ELM has a more complex phase diagram including first- and second-order phase transitions.
In order to apply the method, we need to express the collective spins appearing in the ELM Hamiltonian, Eqs.\ (\ref{eq:Lipkinhamiltonian}) and (\ref{eq:ELipkinhamiltonian}), in terms of individual spins,
\begin{align}
&S_+ = \sum_k a^\dagger_{k,+}a_{k,-} \equiv \sum_k \sigma_{+}^k,  \\ 
&S_-= \sum_k a^\dagger_{k,-}a_{k,+} \equiv \sum_k \sigma_{-}^k, \\
&S_z = \frac{1}{2}\sum_k (a^\dagger_{k,+}a_{k,+} - a^\dagger_{k,-}a_{k,-}) \equiv \frac{1}{2}\sum_k \sigma_{z}^k.
\end{align}
where $\sigma_\pm^k \equiv (\sigma_x^k \pm i\sigma_y^k)/2$.
Here, $\sigma_x^k$, $\sigma_y^k$, and $\sigma_z^k$ are the Pauli matrices for the $k$-th site. $a^\dagger_{k,\pm}$ and $a_{k,\pm}$ are the creation and annihilation operators for a particle in the $k-$th position with spin projection $\pm$ corresponding to the upper or lower levels, respectively. Because the Lipkin model only allows vertical excitations, one can treat the upper and lower states within the same site as independent qubit projections. 
The model can be interpreted in the context of quantum information processing \cite{cervia2021lipkin} without explicitly resorting to the Jordan-Wigner transformation \cite{JordanWigner}, which reduces the size of the Hilbert space by a factor of $2$. In this approach, the standard Lipkin model Hamiltonain is written as,
\begin{equation}
\label{LMG-hamiltonian}
H_L= (1-\lambda) \left( S+\frac{1}{2}\sum_i \sigma_z^i\right)- \frac{\lambda}{2N} \left( \sum_{i<j} (\sigma_{x}^i \sigma_{x}^j - \sigma_{y}^i \sigma_{y}^j)+\sum_{i<j} (\sigma_{x}^i \sigma_{x}^j+\sigma_{y}^i \sigma_{y}^j) \right),
\end{equation}
which leads to the ELM Hamiltonian
\begin{align}
\label{ELM-hamiltonian}
H_{EL} &= H_L - \frac{\lambda}{N} \bigg[\alpha^2 \left(S^2+\frac{1}{4}\sum_{i<j} \sigma_z^i \sigma_z^j+S \sum_{i} \sigma_z^i\right) \nonumber \\
&\quad - \alpha \left( 2S\sum_i \sigma_x^i+\frac{1}{2}\sum_{i<j} \sigma_x^i \sigma_z^j+\frac{1}{2}\sum_{i<j} \sigma_z^i \sigma_x^j\right) \bigg] .
\end{align}
In order to keep the circuits as shallow as possible, we restrict the operator pool to one- and two-body operators that act on particles in at most two pairs of levels. The specific form of the one-body operators that we include in the pool 
are:
\begin{eqnarray}
G_{+}^k &=& \sigma_{+}^k + \sigma_{-}^k = \sigma_x^k,
\label{pool-1}
\\
G_{-}^k &=& -i(\sigma_{+}^k - \sigma_{-}^k) = \sigma_y^k,
\\
G_0^{k} &=& \sigma_z^k,
\end{eqnarray}
while for the two-body operators, with $j<k$, our pool contains the operator structures:
\begin{eqnarray}
T_+^{jk} &=& \sigma_+^j \sigma_+^k + \sigma_-^j \sigma_-^k = \frac{1}{{2}} (\sigma_x^j \sigma_x^k - \sigma_y^j \sigma_y^k),
\\
T_{-}^{jk} &=& -i (\sigma_{+}^j \sigma_{+}^k - \sigma_{-}^j \sigma_{-}^k) = \frac{1}{2} (\sigma_x^j \sigma_y^k + \sigma_y^j \sigma_x^k),
\\
U_+^{jk} &=& \sigma_+^j \sigma_-^k + \sigma_-^j \sigma_+^k = \frac{1}{2} (\sigma_x^j \sigma_x^k + \sigma_y^j \sigma_y^k),
\\
U_-^{jk} &=& -i (\sigma_+^j \sigma_-^k - \sigma_-^j \sigma_+^k) = \frac{1}{2} (\sigma_y^j \sigma_x^k - \sigma_x^j \sigma_y^k),
\\
V_+^{jk} &=& (\sigma_+^j + \sigma_-^j) \sigma_z^k = \sigma_x^j \sigma_z^k,
\\
V_-^{jk} &=& -i (\sigma_+^j - \sigma_-^j) \sigma_z^k = \sigma_y^j \sigma_z^k,
\\
V_0^{jk} &=& \sigma_z^j \sigma_z^k.
\label{pool-10}
\end{eqnarray}
Note that this is the same pool of operators used for the standard Lipkin model in Ref.~\cite{Romero2022}.
\begin{table}[t]
	\centering
	\begin{tabular}{cccccccc}
		\hline
         \hline
 & \multicolumn{2}{c}{\textbf{Path 1}} & \multicolumn{2}{c}{\textbf{Path 2}} &  & \multicolumn{2}{c}{\textbf{Path 3}} \\
& \multicolumn{2}{c}{\textbf{$\alpha=0$}} & \multicolumn{2}{c}{\textbf{$\alpha=\frac{1}{\sqrt{2}}$}} & & \multicolumn{2}{c}{\textbf{$\lambda=1$}} \\
	        \cline{2-3} \cline{4-5} \cline{7-8}
  \textbf{$\lambda$}&  \textbf{Exact}& \textbf{ADAPT-VQE}  &  \textbf{Exact} &\textbf{ADAPT-VQE}&\textbf{$\alpha$} &\textbf{Exact}&\textbf{ADAPT-VQE}\\
  \hline
		0 & 0&0 &0 & 0&0 &-0.9999999 & -0.9999999 \\
  
		0.1 & -0.0184795&-0.0184795 &-0.0189216 &-0.0189216 & 0.1&-1.1051249 & -1.1051249
\\
		0.2 & -0.0437171&-0.0437171&-0.0565697 &-0.0565697 & 0.2&-1.2209975 & -1.2209975
\\		
		0.3 & -0.087747\textbf{4}&-0.087747\textbf{1}&-0.2145\textbf{307} &-0.2145\textbf{288} & 0.3&-1.3483562 & -1.3483562
\\
		0.4 & -0.17137\textbf{33}&-0.17137\textbf{23}&-0.4420\textbf{432} &-0.4420\textbf{389} & 0.4&-1.4879215 & -1.4879215
\\		
		0.5 & -0.288145\textbf{9}&-0.288145\textbf{2}&-0.6892\textbf{442} &-0.6892\textbf{381} & 0.5&-1.6403882 & -1.6403882
\\
		0.6 & -0.420\textbf{1005}&-0.420\textbf{0966}&-0.94511\textbf{39} &-0.94511\textbf{11} & 0.6&-1.8064183 & -1.8064183
\\		
		0.7 & -0.55965\textbf{71}&-0.55965\textbf{21}&-1.20559\textbf{31} &-1.20559\textbf{20} & 0.7&-1.9866367 & -1.9866367
\\
		0.8 & -0.7037\textbf{515} &-0.7037\textbf{458}&-1.46881\textbf{70} &-1.46881\textbf{68} & 0.8&-2.1816263 & -2.1816263
\\
		0.9 & -0.850833\textbf{3}&-0.850833\textbf{0}&-1.7338069 &-1.7338069 & 0.9&-2.3919270 & -2.3919270
\\		
		1   & -0.9999999&-0.9999999&-1.9999999 &-1.9999999 & 1&-2.6180339 & -2.6180339
\\		
\hline
\hline
\end{tabular}
\caption{Exact and simulated ADAPT-VQE ground-state energies per particle for the ELM  with $N=6$ along the three paths described in Section \ref{sec-ELM-paths}, as a function of the corresponding control parameter.}
\label{tab:Exact-Adapt-VQE-ELM-6}
\end{table}

\subsection{Results for the ground state energy}
\label{sec-vqe-results}
In this subsection, we present a comparison of the exact ELM ground state energy values with the approximate results obtained with ADAPT-VQE. 
We employ the same three paths discussed in Section~\ref{sec-ELM-paths}, as they provide a representative sample of the physics at play. 
In addition, we analyze the number of iterations needed for the relative error, $(E_{\rm{ADAPT-VQE}}^{\rm{gs}}-E_{\rm{exact}}^{\rm{gs}})/E_{\rm{exact}}^{\rm{gs}}$, in the ADAPT-VQE calculation to be less than $10^{-5}$, and how this error changes with the number of iterations.
For simplicity, we take as a reference state the lowest energy Slater determinant of the system. We run the ADAPT-VQE simulation for $400$ iterations. Our classical minimization employs the Broyden–Fletcher–Goldfarb–Shanno (BFGS) algorithm 

First, we compare in Table \ref{tab:Exact-Adapt-VQE-ELM-6}
the exact results for the ground state energy per particle for $N=6$ to those obtained with the ADAPT-VQE simulation. The results are presented as a function of the control parameters ($\lambda, \alpha$). The same comparison is also shown in Fig.~\ref{energies-Fig}, where the full orange dots correspond to the ADAPT-VQE results, while the full black line corresponds to the exact ones. Panel a) corresponds to path 1 ($\alpha=0$), panel b) to path 2 ($\alpha=1/\sqrt{2}$), and panel c) to path 3 ($\lambda=1$). It is clear
that the ADAPT-VQE results 
agree closely with the exact results in all the analyzed cases.
The bold figures in Table \ref{tab:Exact-Adapt-VQE-ELM-6} highlight the difference between the exact and the ADAPT-VQE results. 
We find a level of agremeent that in all but one case reaches the fifth significant digit. 
This is remarkable, in the sense that we are exploring a relevant fraction of the model parameter phase space. Note that in path 1 a second order QPT is present in the large-$N$ limit, whereas in path 2 a first order QPT shows up.  
In these two cases, we find some minor differences at the end of our ADAPT-VQE simulation.
\begin{figure}[t]
\centering
\includegraphics[width=0.70\linewidth]{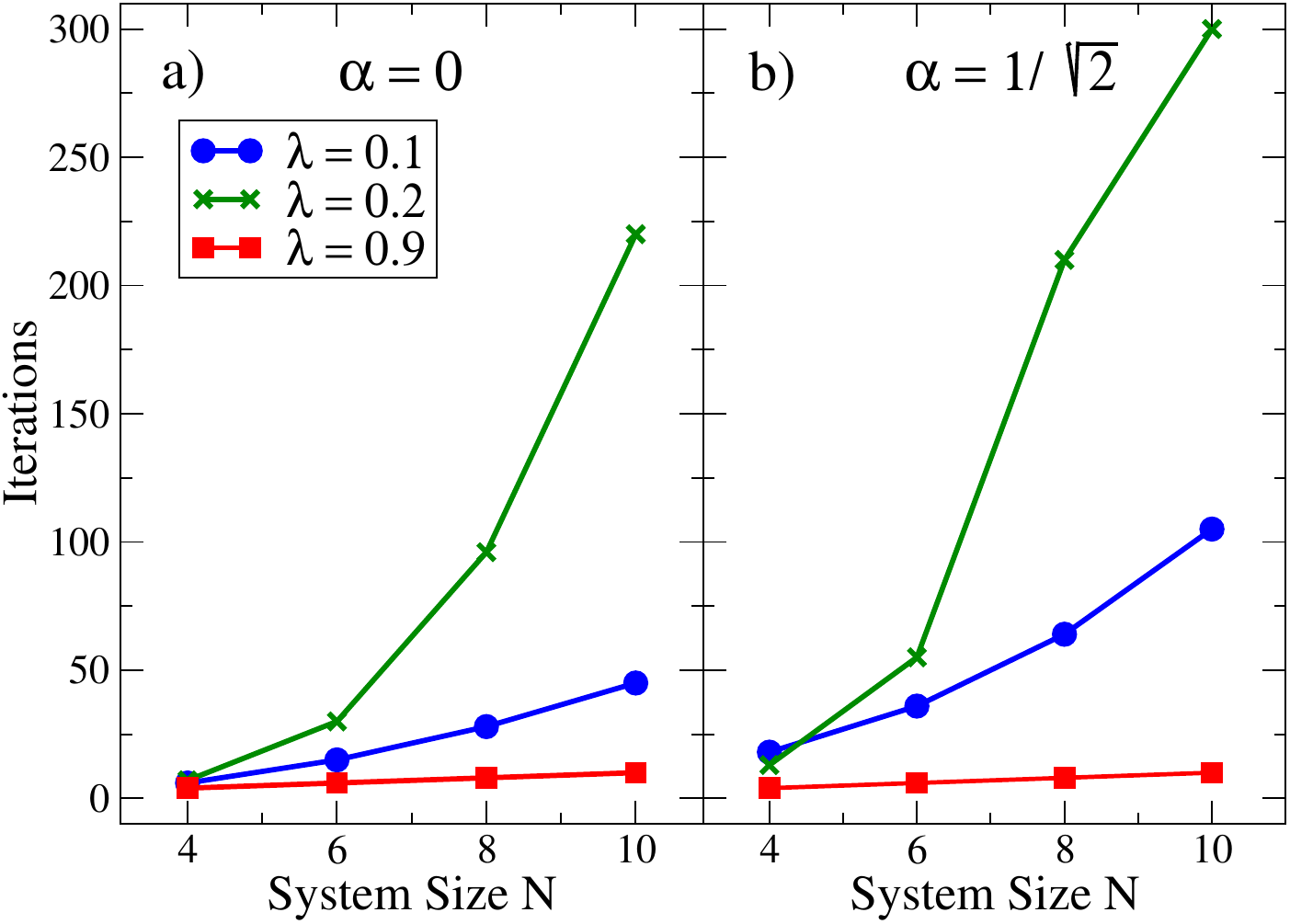}
\caption{ADAPT-VQE number of iterations needed to get quality results (relative errors better than $10^{-5}$) for the ELM ground state energy in path 1 (panel a), and in path 2 (panel b) as a function of the system size $N$. In each panel, three $\lambda-$values are selected: one in the symmetric phase ($\lambda=0.1$), one around the critical line ($\lambda=0.2$), and one in the non-symmetric deformed phase ($\lambda=0.9$). Results for path 3 ($\lambda=1.0$) are very similar to those presented for $\lambda=0.9$ for any value of $\alpha$.}
\label{fig-VQE-convergence}
\end{figure}

In order to complete the information for the ADAPT-VQE calculations, we show in Fig.~\ref{fig-VQE-convergence} the convergence of ADAPT-VQE calculations for the ground state energy of the ELM. Panel a) illustrates the number of iterations required for convergence, set conventionally at a relative difference with the exact value below $10^{-5}$, against the system size $N$ for selected points in path 1, $\alpha=0$. These points include $\lambda=0.1$ (blue line-full dots), $\lambda=0.2$ (green line-crosses) and $\lambda=0.9$ (red line-full squares). Panel b) displays a similar analysis for path 2, $\alpha=1/\sqrt{2}$.
Equivalent figures for path 3, $\lambda=1$, in which the system is always in the non-symmetric phase, are similar to the $\lambda=0.9$ results shown in panels a) and b), independently of the $\alpha$ value.  
From Fig.~\ref{fig-VQE-convergence}, it is clear that the convergence in points corresponding to the non-symmetric phase, $\lambda=0.9$, is reached very fast (red lines) independently of the value of $\alpha$.  
We hypothesize that this is due to the fact that, for all particle numbers, the system has a well-defined deformed minimum. This is true except close to the critical line, which is around the point $\lambda=0.2$ (green lines). Here, there are two competing minima and the convergence of a variational method is expected to be slow. This is indeed confirmed by the corresponding green lines in panels a) and b). In any case, the increase in the number of iterations grows approximately linearly with the size of the system.

\begin{figure}[hbt]
\centering
\includegraphics[width=0.70\linewidth]{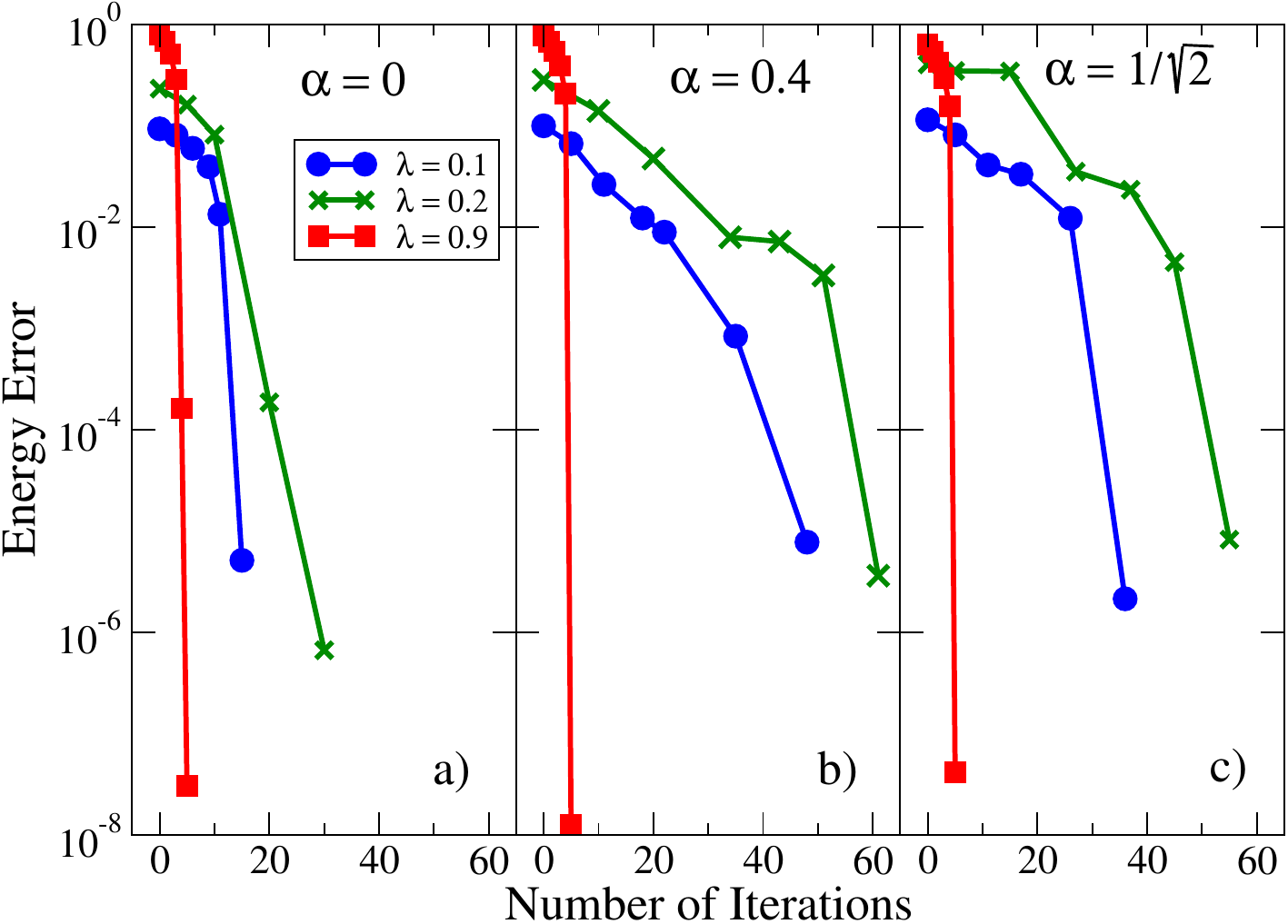}
\caption{Relative ground state energy errors between the ADAPT-VQE calculations and the exact value for the ELM as a function of the iteration number for a system size $N=6$ and for selected points in path 1 [$\alpha=0$, panel a)], path 2 [$\alpha=1/\sqrt{2}$, panel c)] and in an intermediate path [$\alpha=0.4$, panel b)].}
\label{fig-VQE-energy-errors}
\end{figure}

In Fig.~\ref{fig-VQE-energy-errors}, we present the relative errors obtained in the ADAPT-VQE calculations with respect to the exact values of the ELM ground state energies for a system size $N=6$ versus the number of iterations needed. Three selected points in different paths going from symmetric ($\lambda=0.1$) to non-symmetric, well-deformed ($\lambda=0.9$) shapes are presented. Panel a) corresponds to path 1 ($\alpha=0$), panel b) corresponds to an intermediate value of $\alpha=0.4$, and panel c) corresponds to path 2 ($\alpha=1/\sqrt{2}$). 
Again, we find that the faster convergence occurs for $\lambda=0.9$ (red lines), where the system has a well-defined, deformed minimum and ADAPT-VQE finds it easily. The most difficult convergence occurs at the region around the critical line, $\lambda=0.2$, (green lines) where the competition between a spherical and a deformed minimum 
hampers the convergence of the variational method. The symmetric cases, $\lambda=0.1$ (blue lines) are intermediate situations. Regardless of the path, a moderate number of iterations (between $5$ and $60$) is needed to reach convergence in small systems.
We conclude that the ADAPT-VQE with the pool of operators given in (\ref{pool-1}-\ref{pool-10}), as proposed for the standard Lipkin model in Ref.~\cite{Romero2022}, gives an excellent approximation for the ground state energy of the complete parameter space $(\lambda,\alpha)$ of the ELM.

\section{Proposal for the implementation of the ELM model in a digital-analog quantum platform}
\label{Sect-4}
We have so far discussed the ELM and its potential implementation within a VQE. Our findings indicate that ADAPT-VQE can reach the energy minimum even in situations where the system is close to a phase transition in the large-$N$ limit. 
We can alternatively exploit the distinct dynamical features in different regions of the parameter space to identify the different phases of the system~\cite{Perez_Fernandez_2022,Saiz_2022}. This, however, requires access to the temporal evolution of the many-body model at hand.

In this section, we intend to determine the necessary resources for the implementation of the time evolution operator of the ELM Hamiltonian in a quantum simulator. To this end, let us first rewrite the Hamiltonian of Eq.~(\ref{eq:ELipkinhamiltonian}) as
\begin{equation}
    H_{EL} = g_0 + [1-(1+\alpha^2)\lambda] S_z + 2\alpha\lambda S_x - \frac{\alpha^2\lambda}{N} S_z^2 - \frac{4\lambda}{N} S_x^2 + \frac{2\alpha\lambda}{N} (S_xS_z + S_zS_x).
\end{equation}
Considering $S_xS_z + S_zS_x = (S_x + S_z)^2 - S_x^2 - S_z^2$ one gets 
\begin{equation}
    H_{EL} = g_0 + g_z S_z + g_x S_x + g_{zz} S_z^2 + g_{xx} S_x^2 + g_{xz}(S_x+S_z)^2,
    \label{eq:Hamiltonian_Compact}
\end{equation}
where the factors of the collective spin operators have been absorbed into the corresponding $g_j$ coefficients.

Equation~(\ref{eq:Hamiltonian_Compact}) makes apparent what the necessary blocks for the simulation are. The evolution of the operators $S_z$ and $S_x$ can be implemented in the quantum circuit using high-fidelity Single-Qubit Rotation (SQR) gates. Meanwhile, under a purely digital quantum simulation, entangling terms would require a number of lower-fidelity two-qubit (CNOT) gates on top of the SQR gates, which increases with the system size~\cite{Niel2010}. Luckily, in trapped-ion systems, one can efficiently simulate $S_x^2$ by means of a single M{\o}lmer-S{\o}rensen (MS) gate affecting all qubits simultaneously, which evolves analogously to $S_x^2$ \cite{MolmerSorensen,Roos_2008}. Since the error of the two-qubit gates dominates the infidelity of the whole system, this approach offers a significant advantage with regards to purely digital quantum simulations by greatly reducing the required number of two-qubit gates, which is only one, independently of the system size. Note, though, that the fidelity of MS gates in trapped-ions systems still decreases as the number of qubits increases, even if the number of MS gates does not.

\begin{figure}[hbt]
    \centering
    \[
    \begin{array}{c}
    \Qcircuit @C=1.0em @R=0.5em {
    & S_z \ar@{.}[]+<1.2em,1em>;[dd]+<1.2em,-7em> 
    & S_x \ar@{.}[]+<1.25em,1em>;[dd]+<1.25em,-7em>
    & S_x^2 \ar@{.}[]+<1.9em,1em>;[dd]+<1.9em,-7em> 
    &
    & S_z^2
    & \ar@{.}[]+<2.6em,1em>;[dd]+<2.6em,-7em> 
    &
    & (S_x^2 + S_z^2)
    & & \\ 
    & & & & & & & & & & \\
    \lstick{q_1} & \gate{R_z} & \gate{R_x} & \multigate{4}{MS} & \gate{R_y^\dagger(\pi/2)} & \multigate{4}{MS} & \gate{R_y(\pi/2)} & \gate{R_y^\dagger(\pi/4)} & \multigate{4}{MS} & \gate{R_y(\pi/4)} & \qw \\
    \lstick{q_2} & \gate{R_z} & \gate{R_x} & \ghost{MS} & \gate{R_y^\dagger(\pi/2)} & \ghost{MS} & \gate{R_y(\pi/2)} & \gate{R_y^\dagger(\pi/4)} & \ghost{MS} & \gate{R_y(\pi/4)} & \qw \\
    & \vdots & \vdots & & \vdots & & \vdots & \vdots & & \vdots & \\
    &   & & & & & & & & & \\
    \lstick{q_n} & \gate{R_z} & \gate{R_x} & \ghost{MS} & \gate{R_y^\dagger(\pi/2)} & \ghost{MS} & \gate{R_y(\pi/2)} & \gate{R_y^\dagger(\pi/4)} & \ghost{MS} & \gate{R_y(\pi/4)} & \qw 
    }
    \end{array}
    \]
    \caption{Quantum circuit implementing the evolution operator of the ELM Hamiltonian of Eq.~(\ref{eq:Hamiltonian_Compact}). Here, $R_\nu$ are single qubit rotations in the $\nu \in {x,y,z}$ axis and $MS$ is a single M{\o}lmer-S{\o}rensen gate affecting all qubits, which acts as the analog block for this implementation.}
    \label{fig:Circuit}
\end{figure}
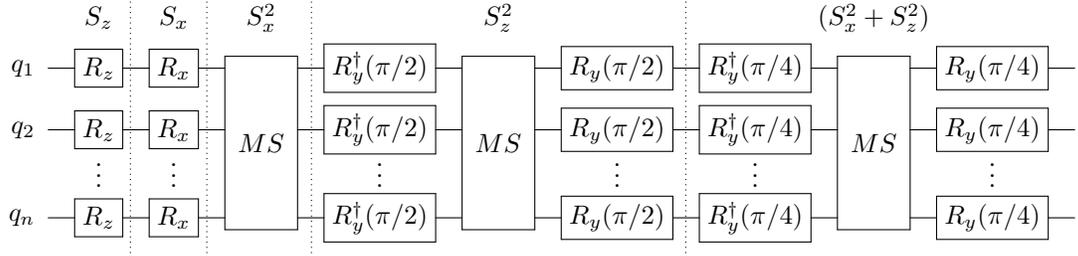

Considering the convenience of these MS gates, we propose an implementation based on the recently developed Digital-Analog Quantum Computation (DAQC) paradigm~\cite{Celeri_2023,DAQS_Review}. By means of employing analog blocks as the entanglement sources of the system, alongside high-fidelity SQR gates, one can achieve universal quantum computation, exploiting both the robustness of Analog Quantum Simulations and the versatility of Digital Quantum Computers. Fig.~\ref{fig:Circuit} shows a scheme of the proposed quantum circuit.

First, the evolution of a single body term $G_\nu = g_\nu \sum_i^N \sigma_\nu^i$, where $\sigma_\nu^i$ is the Pauli operator $\nu \in {x,y,z}$ acting upon qubit $i$, can easily be implemented with a SQR gate $R_\nu^i(\theta) = \exp{[i\frac{\theta}{2}\sigma_\nu^i]}$, with $\theta = g_\nu t$, applied to every qubit. So, the terms $S_x$ and $S_z$ only require $N$ single-qubit gates each.

Second, as previously mentioned, making use of the properties of trapped-ion systems, the term $S_x^2$ can be directly implemented as an analog block via a single MS gate, which is independent of $N$. The MS gate consists of a bichromatic laser field acting on the qubits, and the parameter can be controlled through the phase and frequency of this laser field. Lastly, the terms $S_z^2$ and $(S_x + S_z)^2$ can be implemented in a similar manner, by applying SQRs $R_y(\pi/2)$ and $R_y(\pi/4)$, respectively, to all qubits before and after the MS gate. Effectively, this performs a local rotation from the $z$ or $x+z$ axes to the $x$ axis, which can then be evolved analogically with a MS gate, then rotated back to its original axis. This means that all three two-body terms can be implemented using only three two-qubit MS gates (one for each term), plus $4N$ single-qubit gates to rotate $S_z^2$ and $(S_x + S_z)^2$ into the $x$ direction and then back ($2N$ gates each).

Therefore, for each single-body term that is implemented, we need $N$ gates, for a total of $2N$ (there are two single body terms). For each two-body term that we need to rotate to the $x$ axis, we also need $2N$ single-qubit gates. Because we have three two-body terms, but one of them already points along the $x$ axis, we only need $4N$ single-qubit gates. The total number of gates for the circuit would then be $6N$ single-qubit gates plus three MS gates. However, since not all the terms in the Hamiltonian commute with each other, applying these gates sequentially results in an evolution that differs from that of the target Extended Lipkin Hamiltonian. A common way to deal with this error is to employ the first-order Lie-Trotter-Suzuki formula \cite{trotter1959product,georgescu2014quantum}:
\begin{equation}
    U(t) \approx \left( \prod_k e^{-i H_k t / n_T} \right)^{n_T} = U_T(t,n_T),
\end{equation}
where $n_T$ are the so-called Trotter steps and $H_k$ are the $k$ non-commuting terms that conform the Hamiltonian, i.e, $H = \sum_k H_k$. For this particular case, the terms are chosen as follows:
\begin{align}
    H_1 =& ~ g_zS_z + g_{zz}S_z^2, \\
    H_2 =& ~ g_xS_x + g_{xx}S_x^2, \\
    H_3 =& ~ g_{xz} (S_x + S_z)^2.
\end{align}
One can approximate the exact evolution by performing the evolution in time $t$ in many small steps \mbox{$n_T = t/\Delta t$} of time $\Delta t$, called Trotter steps. 
The approximation becomes exact in the limit $n_T \rightarrow \infty$ $(\Delta t \rightarrow 0)$, which means that it can be made arbitrarily precise by increasing the number of Trotter steps. Note, though, that, in order to do so, one must increase the depth of the circuit, aggregating the infidelities of the quantum gates in the circuit for each run. Therefore, it is important to study, case by case, the experimental error introduced with each Trotter step versus the digital one due to the Trotter approximation. 

Fidelities of up to $99.9999\%$ have been reported for single-qubit gates~\cite{singlequbitfidelity}, while fidelities of up to $99.9\%$ have been achieved for MS gates in trapped-ion systems~\cite{twoqubitfidelity}.
Since the infidelities of individual gates are small enough and given the resource quantification we just presented, it is reasonable to assume that the total fidelity of the system will approximately be
\begin{equation}
    f = \left[ (f_s)^{6N}(f_{MS})^3 \right] ^{n_T},
    \label{eq:gate_fidelity}
\end{equation}
where $f_s$ and $f_{MS}$ are the fidelities of single-qubit and MS gates, respectively.
\begin{figure}[hbt]
    \centering
    \hspace{-15pt}
    \begin{tabular}{cc}
        \textbf{a)} & \textbf{b)} \\
    \includegraphics[width=0.52\textwidth]{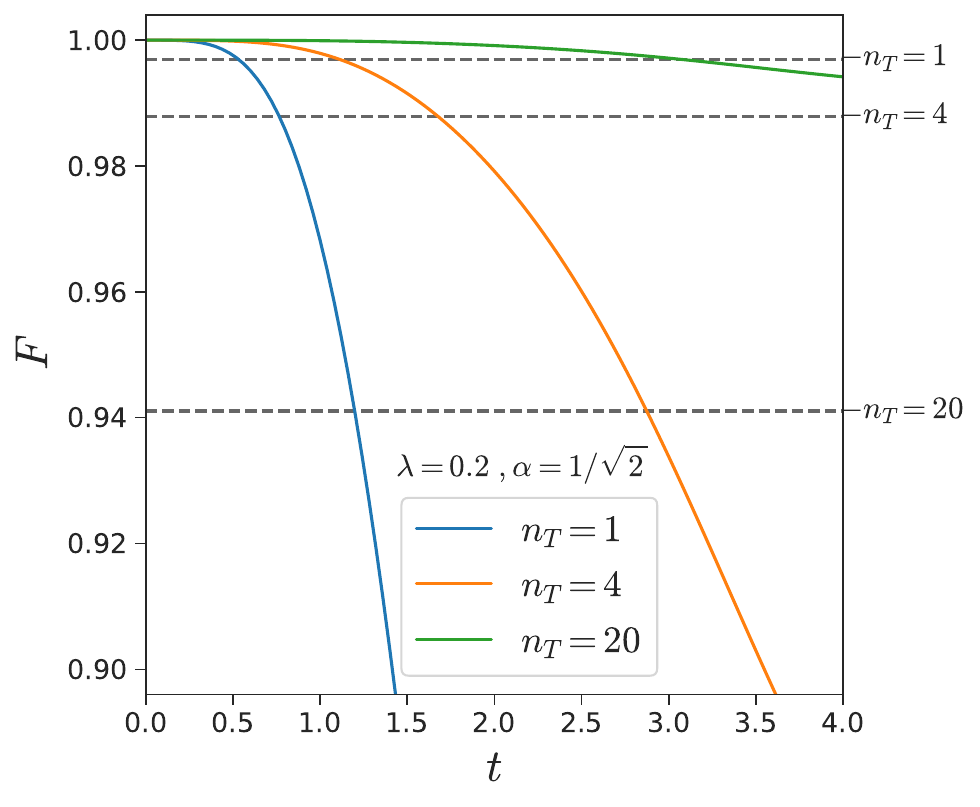} & 
    \includegraphics[width=0.47\textwidth]{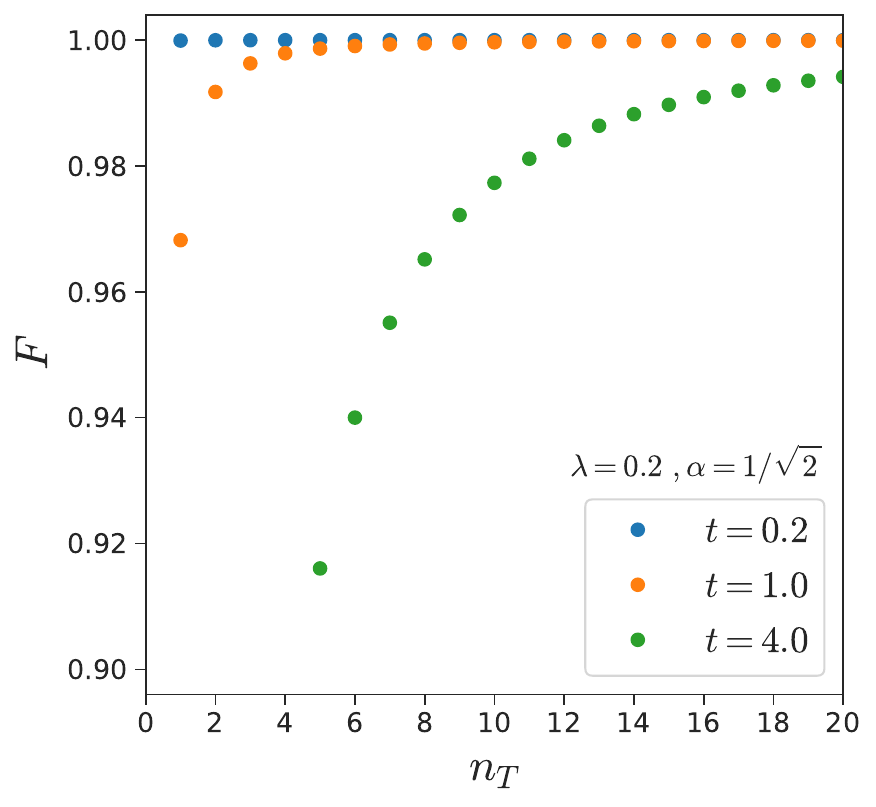}
    \end{tabular}
    \caption{ Fidelity of the trotterized evolution according to Eq.~(\ref{eq:trotter_fidelity}). The system size is $N=6$ and the parameters $\lambda = 0.2$, $\alpha= 1/ \sqrt{2}$ in all cases. The initial state is $| \Phi_0 \rangle = |\uparrow \downarrow \uparrow \downarrow \uparrow \downarrow \ \rangle $. Panel a): fidelity vs evolution time for several numbers of Trotter steps. Horizontal dashed lines indicate total circuit gate fidelity as per Eq.~(\ref{eq:gate_fidelity}), for a number of Trotter steps. Panel b): fidelity vs number of Trotter steps for several values of the evolution time.}
    \label{fig:fidelity}
\end{figure}

One can define the fidelity of the Trotter approximation for a particular initial state $|\phi(0)\rangle$ as
\begin{equation}
    F = |\langle \phi (0) | U_T(t,n_T) U(t) | \phi(0) \rangle|^2.
    \label{eq:trotter_fidelity}
\end{equation}
The values of the fidelity of the Trotter evolution are shown in Fig.~\ref{fig:fidelity} for different time values and Trotter steps in a system of size $N=6$. In panel a), we find that, for a low number of Trotter steps, the fidelity decreases at a very fast rate with time. This suggests that the digital error will dominate over the gate error, except for short or very short times. In contrast, a high number of Trotter steps will allow longer evolution times without incurring strong infidelities due to digital errors, but with a high gate infidelity that will dominate the total error. Similarly, in panel b) of Fig.~\ref{fig:fidelity}, we show how the fidelity improves with $n_T$ for different values of time. For a given amount of evolution time $t$, there is a value of $n_T$ beyond which there is no substantial improvement of the fidelity.

This leads to a compromise in fidelity, as decreasing the errors from one source implies increasing the errors from the other. This often requires finding a balance between the precision of the Trotter expansion and the precision of the physical implementation, but, as it will be shown in the following section, for certain tasks, such as finding the phase diagram of a quantum system, ML approaches can properly work with a reduced number of Trotter steps, saving computational resources and thus minimizing the total gate infidelity.

\section{Unveiling the ELM phase diagram through a Machine Learning analysis}
\label{Sect-5}

We now discuss specifically an ML implementation to identify the different phases of the ELM.
In Section \ref{sec-ph-diagram}, we have discussed the use of the expectation value of $\langle S_z \rangle + N/2$ for the ground state as an order parameter. In Fig.~\ref{fig-ELM-order-param}, one can clearly observe the precursors of the QPT for a relatively small system of size $N=6$, far from the large-$N$ limit. In this section, we will present a different way of identifying the critical points of the QPT. We will employ ML techniques that can enhance our ability to precisely classify the quantum phase of the system. First, we will exploit our previous theoretical knowledge of the phase diagram using supervised ML algorithms; then, we will propose strategies to gather new information in systems where the phase diagram is unknown with the aid of unsupervised ML methods.

\subsection{Supervised quantum phase classification}
\label{sec-supervised}
A common way of identifying a QPT is by measuring an order parameter, which 
for the ELM Hamiltonian can be $(\langle S_z \rangle + N/2)/N$ (see  Fig.~\ref{fig-ELM-order-param}). Nevertheless, in practice, this implies  \emph{a priori} knowledge of the ground state of the system, which may be a big challenge in certain many-body systems e.g., in the Heisenberg or the Ising models. Therefore, other less costly methods may be of interest, especially if it is not necessary to obtain the ground state of the system. 
In Refs.\ \cite{Saiz_2022, Perez_Fernandez_2022}, some of us proposed to use the time evolution of an appropriately chosen operator as a proxy of the phase of the system. In particular, the following time-dependent function was used 
\begin{equation}
    C_\nu(i,j,t) = \langle \Phi_0 |U(t)^{\dagger} \sigma_\nu^i \sigma_\nu^j U(t)| \Phi_0\rangle - \langle\Phi_0| U(t)^{\dagger}\sigma_\nu^i U(t)| \Phi_0\rangle \langle \Phi_0|U(t)^{\dagger}\sigma_\nu^j U(t)|\Phi_0 \rangle,
    \label{eq:correlation}
\end{equation}
where $\sigma_\nu^i$ corresponds to the Pauli matrix on the axis $\nu = \{ x,y,z \}$ over the $i$-th qubit and $U(t)$ to the evolution operator. Note that this function depends on the considered state, $|\Phi_0\rangle$, that should not be an eigenstate of the system. 
We will also employ  the time evolution of the average value of $S_z$,
\begin{equation}
S_z(t) = \langle \Phi_0 | U(t)^{\dagger} S_z U(t) | \Phi_0 \rangle,  
\label{Szt}
\end{equation}
which we intuitively expect to be sensitive to the phase of the system. 
However, inferring the phase is not always a trivial task. 
Among other things, one needs to find a rule that connects the time evolution of these observables with the phase of the system.

\begin{figure}[t]
    \centering
    \begin{tabular}{cc}
        \begin{minipage}{0.45\textwidth} 
            \includegraphics[width=0.9\textwidth]{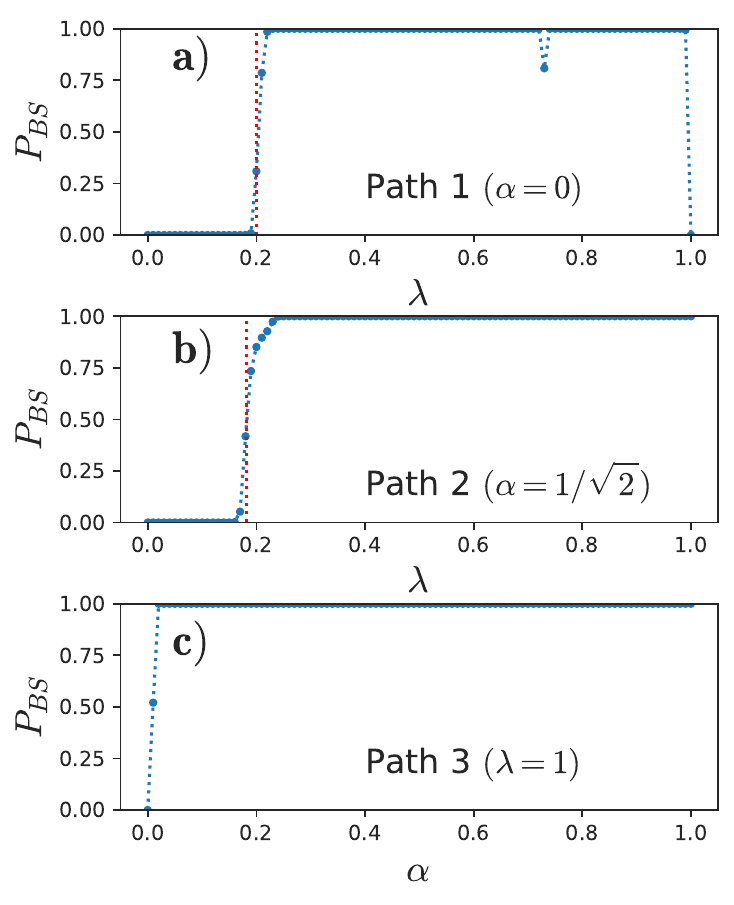}
        \end{minipage} &
        \begin{minipage}{0.55\textwidth} 
            \includegraphics[width=1\textwidth]{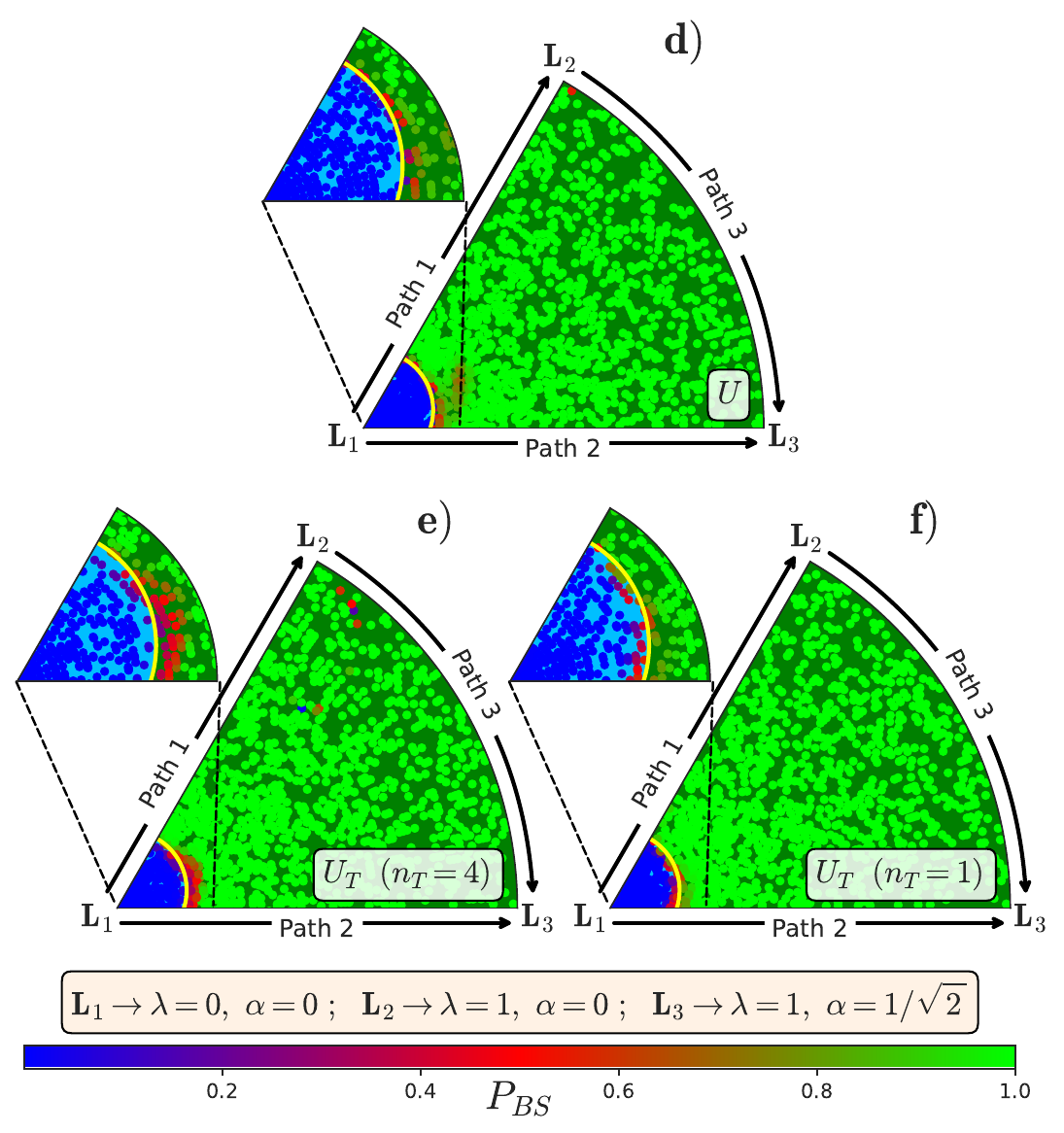}
        \end{minipage}
    \end{tabular}
    \caption{Phase probabilities, $P_{BS}$, for the testing dataset as predicted by the CNN using $S_z (t)$, 
    $N=6$, and an initial state $| \Phi_0 \rangle = 1/\sqrt{3}(|m_z=-N/2\rangle + |m_z=0\rangle + |m_z=N/2\rangle )$. Panels a), b), and c): $P_{BS}$ for three relevant paths (see text) using the exact evolution operator for training and testing. Panel d): $P_{BS}$ using the exact evolution operator ($U$) for training and testing. 
    Panel e): $P_{BS}$ using the Trotter evolution operator ($U_T$) with $n_T=4$ for training and testing. Panel f): $P_{BS}$ using the Trotter evolution operator ($U_T$) with $n_T=1$ for training and testing. Each point corresponds to a single evolution of the test dataset. Light blue and dark green backgrounds fill the phase diagram for the symmetric and the broken symmetry phases, respectively, and a yellow line denotes the critical line between both phases.}
    \label{fig:predictionsSz}
\end{figure}

A very appealing way to establish the relationship between the phase and the time evolution is to use supervised ML based on our knowledge of the phase diagram (see Fig.~\ref{phaseD}). For example, one could use labeled data to train a Neural Network as a classifier, which would then be able to recognize the phase in which the system is in, based only on the measurement of its order parameter \cite{lecun2015deep}. 
Furthermore, ML allows for the use of less intuitive or visually apparent methods. This expands the set of tools available to study QPTs, using different operators as probes of the QPT or even initial states that are not the ground state, which is our objective.

There is no single ML algorithm that consistently performs classification tasks better than any other algorithm \cite{Schmidhuber_2015}. In other words, multiple different approaches can be equally viable. In our specific case, we choose to build and train a Convolutional Neural Network (CNN) \cite{CNN701181}. This choice is motivated by the well-known success of CNNs in recognizing shapes and patterns as image classifiers and other visual recognition tasks \cite{CNNTraffic,Lin_2023}, as well as by its application in previous similar works \cite{Saiz_2022}. The CNN consists of a standard architecture: $3$ Convolution-Relu-Pooling layers with $32$ kernels of size $3$, plus a set of $4$ fully-connected (dense) layers consisting of $512$ neurons each. We also introduce dropout layers as a measure against data overfitting. 

To train the CNN, a dataset is needed. Since the ELM is exactly solvable, the dataset is created by classically computing the evolution of the expected value of the desired operators, of Eqs.~(\ref{Szt}) or (\ref{eq:correlation}),
for a system with $N=6$ particles.
These dynamical evolutions are computed for values of the control parameters $\lambda \in [0,1]$ and $\alpha \in [0,1]$, in $0.01$ steps, yielding a grid of $10201$ data points. Each evolution is computed for a conveniently adimensionalized time $t \in [0,4]$ with steps of size $0.04$. 
Then, we label our simulations as either 0, if they correspond to the symmetric phase, or 1, if they correspond to the broken-symmetry phase. 
We remove the simulations of the 
points corresponding to the three paths mentioned in Sect.\ \ref{sec-ph-diagram} from the dataset. These, alongside a randomly selected 20\% of the remaining points, are kept apart for testing purposes. The remaining data is then used for training and validation of the training. The cross-validation strategy employs a randomly selected 20\% of the non-test data at each iteration (epoch) of the training. This validation data is used to confirm that there is no overfitting, sudden forgetting or other anomalies during the training process, as well as to optimize the hyperparameters of the CNN, such as the number of layers and the number of neurons per layer. The test dataset is used exclusively to compute the results reported here.  

First, we use $S_z(t)$ in Eq.~\eqref{Szt} to train the CNN with the labeled data. The output of the network is not a discrete classification; instead, it provides the probability, $P_{BS}$, 
for the corresponding point in the phase diagram to belong to the broken symmetry (BS) phase. Ideally, this probability should be close to $0$ in the symmetric phase, close to $1$ in the broken phase, and in between those values when close to a critical point. 
Once the CNN has been trained, the performance is analyzed with the 
dataset that was not considered in the training. 

Figure~\ref{fig:predictionsSz} depicts the probability $P_{BS}$ for the testing dataset using the evolution of the operator $S_z (t)$, a system size $N=6$, an initial state $|\Phi_0 \rangle = 1/\sqrt{3}(|m_z=-N/2\rangle + |m_z=0\rangle + |m_z=N/2\rangle )$ and control parameters $\alpha \in [0,1/\sqrt{2}]$ and $\lambda \in [0,1]$. We show in panels a), b), and c) the results for the three relevant paths, considering the exact evolution for training and predicting. Other than the inaccuracy at the extremes of paths 1 and 3, for values $\lambda=1$ and $\alpha=0$, respectively, the rest of the predictions are precise and mostly with complete certainty. Paths 1 and 2 show a very accurate prediction of the critical point (marked with the red dotted line), as $P_{BS}$ suddenly jumps from 0 to 1 precisely in this area, despite working with a relatively small system size. Panels d), e) and f) (including the insets which provide a zoomed view) give a more general view of the quantum phase diagram classification (using also $S_z (t)$ and $N=6$) across the entire parameter space, which is, once again, very precise. In panel d), the system is trained with the exact evolution operator ($U$). Symmetric and broken phases are well determined and the critical line (yellow line) is nicely reproduced, although a small part of the BS phase is predicted as symmetric (with low certainty). In panels e) and f), the Trotter evolution is considered, with $n_T=4$ and $n_T=1$, respectively. It is remarkable that these results are of the same quality as the exact ones, specially for $n_T=1$.  This fact is particularly relevant as it means that it is not necessary to correct the Trotter infidelities, saving computational resources and, thus, reducing the total infidelity incurred by lowering the number of required quantum gates.

\begin{figure}
    \centering
    \begin{tabular}{cc}
        \begin{minipage}{0.45\textwidth} 
            \includegraphics[width=0.9\textwidth]{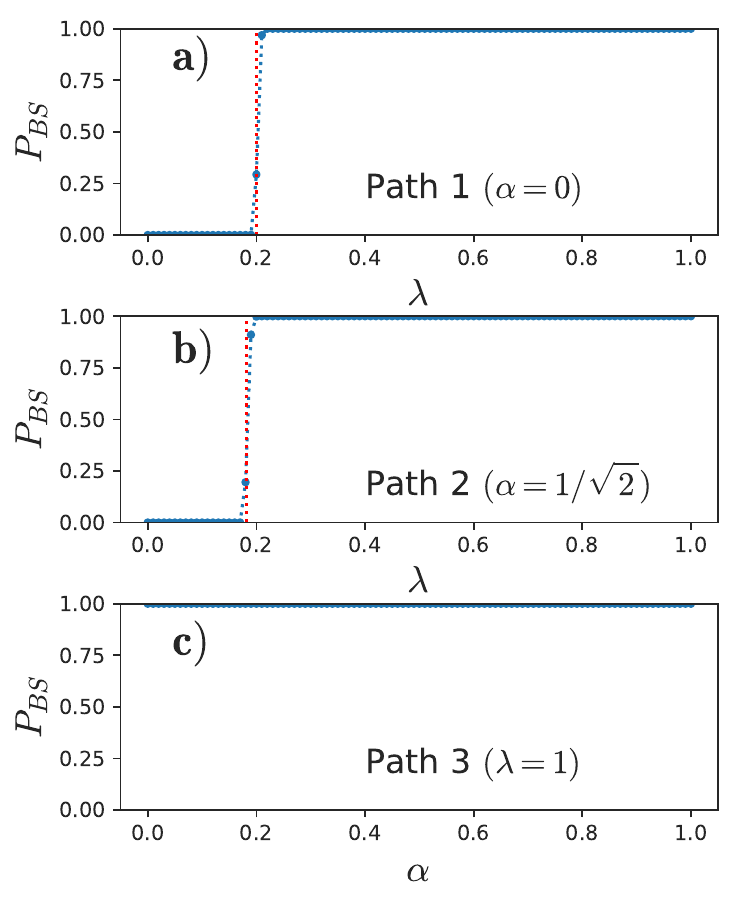}
        \end{minipage} &
         \begin{minipage}{0.55\textwidth} 
        \includegraphics[width=1\textwidth]{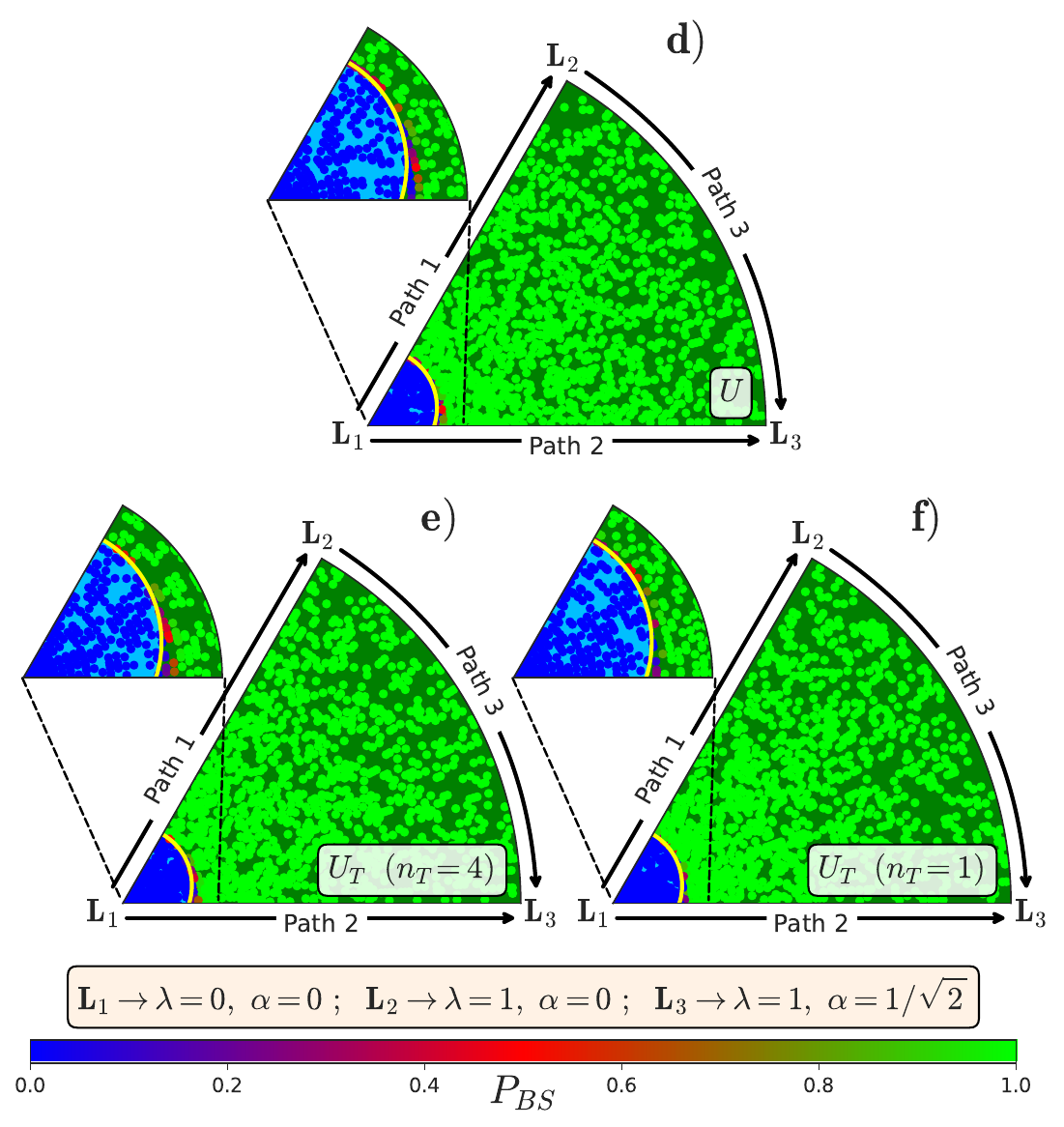}
        \end{minipage}

    \end{tabular}
    \caption{Same as Fig.~\ref{fig:predictionsSz}, but using the function $C_z(1,2,t)$ computed in the spin basis for a system of size $N=6$, with initial state $| \Phi_0 \rangle = |\uparrow \downarrow \uparrow \downarrow \uparrow \downarrow \ \rangle $.}
    \label{fig:predictionsCz12}
\end{figure}

Second, we use the correlation operator of Eq.~(\ref{eq:correlation}) to train the CNN to predict the phase diagram. This option is motivated by the results obtained in Refs.\ \cite{Perez_Fernandez_2022,Saiz_2022}.
Choosing different correlation functions can yield slightly different results and, due to the nature of deep learning algorithms, it is usually not possible to intuitively guess which one may be the best option. We stress that, while the following results were achieved using the  $C_z(1,2,t)$ correlation function, analogous results were achieved with all possible combinations of $i,j$. The correlations $C_y(i,j,t)$ and $C_x(i,j,t)$ might also be viable probes. We follow the same procedure just described for $S_z$, replacing its evolution for the time evolution of $C_z(1,2,t)$ as input of the CNN. In addition, the initial state $|\Phi_0 \rangle = | \uparrow \downarrow \uparrow \downarrow \uparrow \downarrow \ \rangle $ was arbitrarily chosen. Note that, since $\sigma_z^i$ acts on a particular qubit, one cannot compute the correlation function in the collective spin basis. 

The results for this second choice of correlator are shown in Fig.~\ref{fig:predictionsCz12}.
We reach the same conclusions as for $S_z(t)$, but with an improvement in the ability of the CNN to correctly classify the quantum phase of the system and precisely identify the critical point. For the three selected paths, the CNN only shows uncertainty for the points closest to the critical points in paths 1 and 2 ($\lambda = 0.20$ and $\lambda = 0.18$), which is the best possible guess. 

This ML approach to phase classification has another advantage, which is its robustness against errors introduced before the training. This is generally true for properly constructed CNNs \cite{GHAFARI2021107009}, but they become particularly robust for data such as the one at hand. The evolution of $C_z(1,2,t)$ at each $(\lambda,\alpha)$ value has a particular shape, and the CNN learns the features of these shapes to classify them. The variation of the parameter $\lambda$ has a very prominent effect on the shape of this evolution, while the parameter $\alpha$ has a relatively small effect in comparison. With the addition of  noise, data points at different values of $\alpha$ for the same value of $\lambda$ become difficult to distinguish, but the CNN is still able to, overall, differentiate points at different values of $\lambda$. Moreover, the QPT depends mostly on $\lambda$, so even if the details on $\alpha$ are lost, the CNN can still classify most of the phases correctly, albeit with a shift in the exact location of the critical line. 

\begin{figure}[hbt]
    \centering
    \begin{tabular}{cc}
    \textbf{a)} & \textbf{b)} \\
            \hspace{-15pt}
            \includegraphics[width=0.30\textwidth, trim={0 0 0 0},clip]{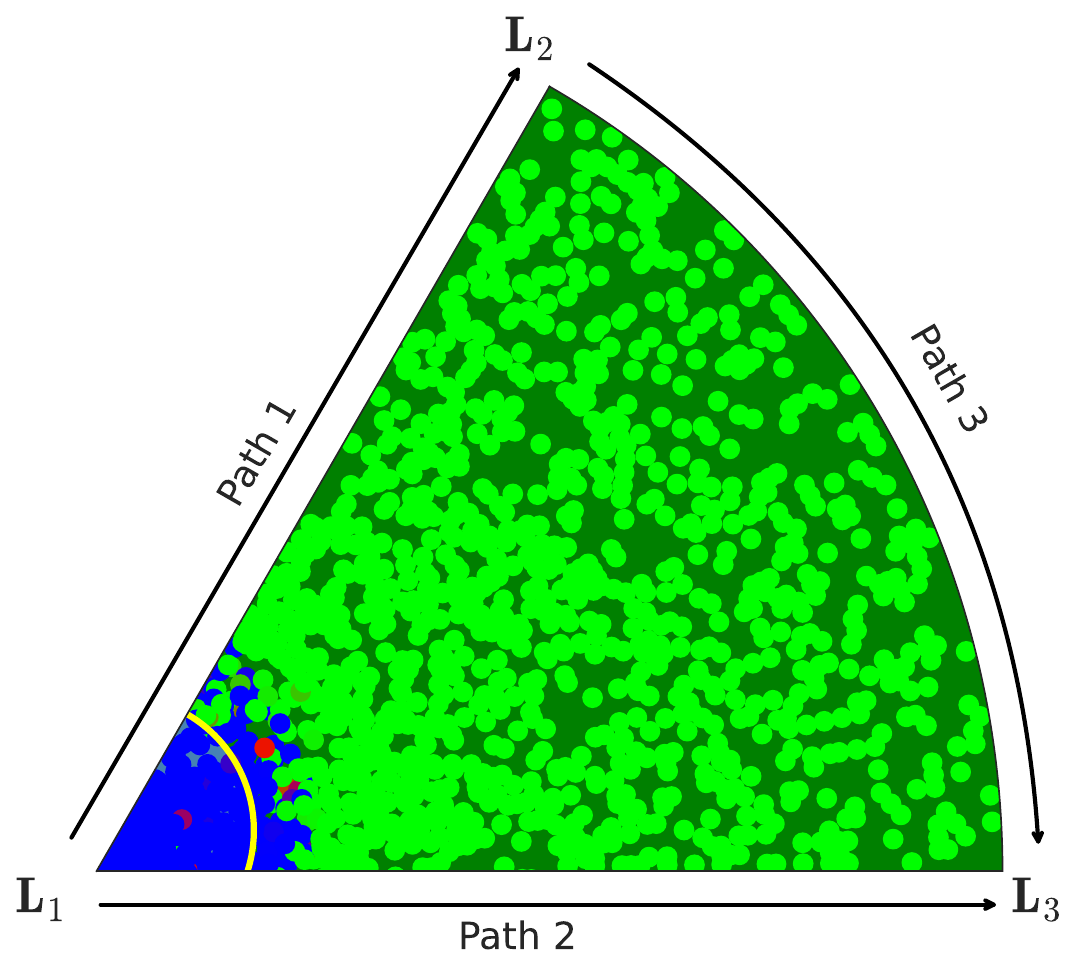} &
           \includegraphics[width=0.30\textwidth, trim={0 0 0 0},clip]{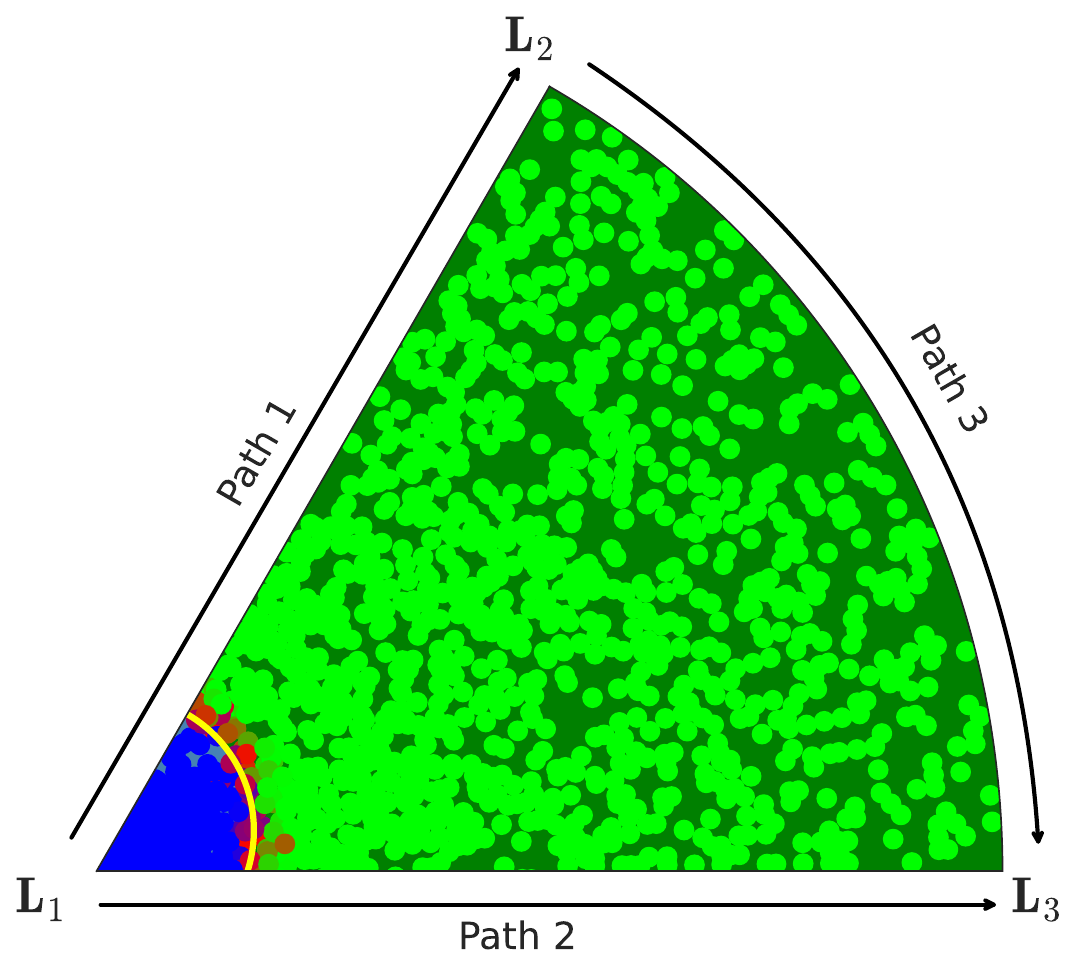}
           \hspace{0pt}
           \end{tabular}
           \hspace{-15pt}
           \includegraphics[width=0.56\textwidth]{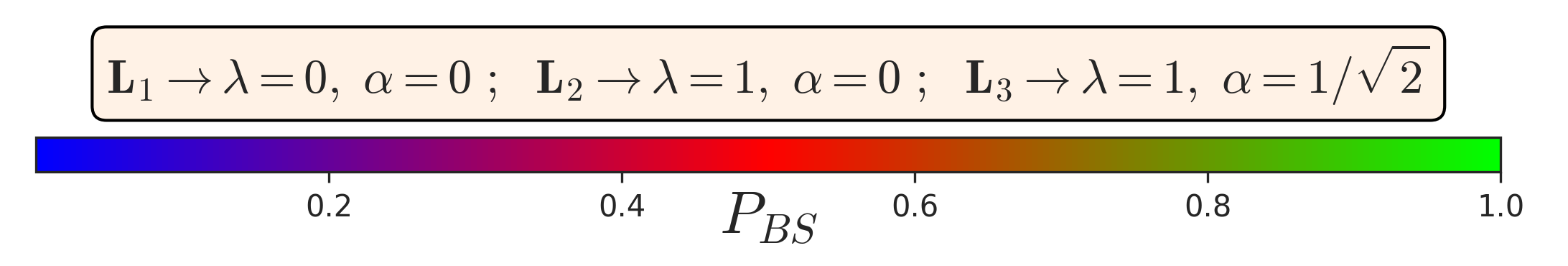}
    \caption{Effect of noise. Phase probabilities, $P_{BS}$, for the testing dataset as predicted by the CNN using $C_z(1,2,t)$, a system size $N=6$, an initial state $| \Phi_0 \rangle = | \uparrow \downarrow \uparrow \downarrow \uparrow \downarrow \ \rangle $ and using the trotterized evolution $U_T(n_T=1)$. Panel a): results with a CNN trained on noiseless data and tested on noisy data (see text). Panel b): results with a CNN trained and tested on noisy data (see text).} 
    \label{fig:predictionsCz12_noiseless_noise}
\end{figure}

Having said that, one can train the Neural Network with the same kind of noisy data that one is expecting to predict. This allows the CNN to, in a way, learn the noise pattern, recovering a precise prediction of the QPT. Both of these issues are shown in the predictions of Fig.~\ref{fig:predictionsCz12_noiseless_noise}, where a random Gaussian noise with a standard deviation $\sigma = 0.2$ was added to the input data. Panel a) shows the results for a CNN trained with noiseless data, when predicting the phases of noisy data. We find that the predicted phase transition shifts, and the predictions of the symmetric phase enter the area of the BS phase, but the overall phase classification remains quite accurate. In contrast, panel b) shows a CNN trained on noisy data. Here, the misclassification issue is fixed and the phase transition is once again accurately predicted. We note, however, that infidelities in quantum gates  and other sources of errors in quantum simulators are biased, unlike random noise. It remains unclear whether the infidelities of a real quantum simulator would affect the predictions of the CNN, although it is safe to assume that significantly smaller errors will have a significantly smaller impact.

\subsection{Supervised classification with partial information}
\label{sec-partial}
The main and most obvious disadvantage of supervised ML algorithms is that they require computing or labeling the data with prior knowledge of the correct classification. For that reason, one of the main goals of many supervised ML algorithms is to achieve a good enough generalization using a reduced dataset. 
We explore the generalization ability of our setup as follows.
First, we train our CNN only with data corresponding to the Lipkin model of Eq.~(\ref{eq:Lipkinhamiltonian}), that is, with $\alpha=0$ in the ELM Hamiltonian (\ref{eq:ELipkinhamiltonian}). Then, we try to predict the quantum phase diagram for the ELM, with $\alpha \neq 0$. Note that the regular Lipkin Hamiltonian exhibits only a second-order QPT, while the ELM presents a first-order QPT for values of $\alpha \neq 0$ \cite{Vida2006}.

\begin{figure}[hbt]
    \centering
        \hspace{-15pt}
        \includegraphics[width=0.45\textwidth, trim={0 0 0 0},clip]{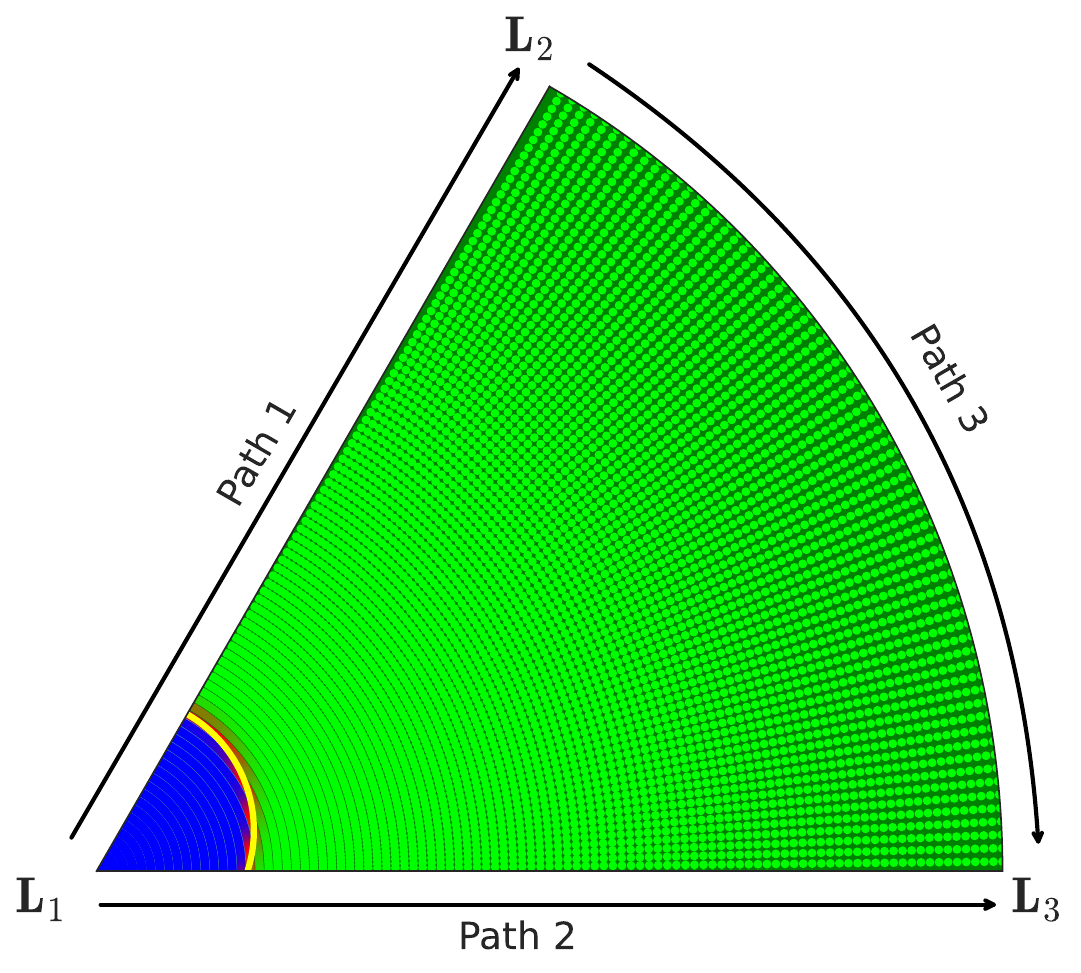} \\
       \includegraphics[width=0.5\textwidth]{Figures/Colorbar.png}
    \caption{ The same as in Fig.~\ref{fig:predictionsCz12}, but training the CNN only with points within path 1 ($\overline{L_1 L_2}$), i.e. $\alpha = 0$.}
    \label{fig:predictions_alphazero}
\end{figure}

Figure~\ref{fig:predictions_alphazero} shows the predicted quantum phase diagram using this approach. Since the CNN had no information about points with $\alpha\neq 0 $, we may expect  the predictions to become inaccurate as we move towards large values of $\alpha$. However, the CNN predictions remain surprisingly accurate for large values of $\alpha$. We take this as a sign of the generalization capabilities of CNN's. We stress that these results are not trivial. We are able to predict a first-order QPT in a fundamentally different Hamiltonian, with only data from the second-order QPT of the known Lipkin model. The prediction with only partial information has a slightly smaller categorical accuracy ($98.86\%$) when predicting the quantum phase of the system, compared to the results of Fig.~\ref{fig:predictionsCz12} ($99.85\%$). In other words, using the complete quantum phase space for training provides marginally more accurate predictions. The difference is, however, not easily observed in Fig.~\ref{fig:predictions_alphazero}. 

\subsection{Unsupervised classification}
\label{sec-unsuper}

Finally, we are interested in correctly predicting the phase diagram of the system without any prior theoretical knowledge about it. To achieve this, the use of unsupervised methods, which do not employ labeled training data, becomes necessary. Of course, with this approach, new challenges arise. The lack of labeled data for training means that these algorithms rely on the definition of some sort of similarity between the data.  

As will be shown below, unsupervised ML will allow to distinguish points near the critical area from those that lay well apart. This will provide a way to deduce the quantum phase diagram, but requires some previous knowledge of the model at hand, i.e., one needs to know where different phases exist. Of course, for more complex quantum phase diagrams with more than a single QPT, a different approach is likely to be necessary.

There are many possible unsupervised ML algorithms that could be used, namely, clustering, factor analysis, principal component analysis, and embedding techniques.  
Analogously, different observables have been studied as input for the unsupervised algorithms, concluding that the evolution of the functions $C_z(1,2,t)$ and $S_z(t)$, which were used in the supervised ML, are unsuitable for the unsupervised cases, being impossible to obtain conclusive results. This was expected as part of the limitations of unsupervised ML, which cannot easily learn specific patterns or features to classify. 


We consider a different alternative and use as input the different correlation functions that arise from all possible combinations of $i,j$, and $\nu$ in Eq.\ (\ref{eq:correlation}). This means that, in a system of size $N=6$, for a single data point, i.e. a single value of $\alpha$ and $\lambda$ for the Hamiltonian, $45$ correlation functions are computed (note that $C_\nu(i,i) = 0$ and $C_\nu(i,j) = C_\nu(j,i)$). We use these correlations for a clustering algorithm known as Fuzzy C-means \cite{fuzzyCmeansBezdek,fuzzyCmeansDunn}, which is a variation of the popular K-means clustering algorithm \cite{Lloy1982}. 

K-means \cite{Lloy1982} is a general-purpose clustering algorithm in which, given a set of $n$ samples $\vec{X} = \{ \vec{x}_i, \dots, \vec{x}_n \}$ and a number of clusters $K$, cluster centroids $\vec{C} = \{ \vec{c}_1, \dots, \vec{c}_K \}$ are computed to minimize the  objective function,
\begin{equation}
    f(\vec{C}) = \sum_i^n \sum_j^k w_{i,j} (|| \vec{x}_i - \vec{c}_j ||)^2,
    \label{eq:partition}
\end{equation}
where $w_{i,j} \in \{ 0, 1 \}$ is a partition matrix that tells us whether sample $i$ belongs to cluster $j$ ($w_{i,k}=1$) or not ($w_{i,k}=0$). We implicitly assume that $\vec{x}_i$ and $\vec{c}_j$ correspond to multidimensional vectors.
Once the number of clusters has been fixed, cluster centroids are initialized, commonly either randomly or through some initialization function that may speed up the convergence, and each sample is assigned to the closest centroid. Then, the mean position of the samples assigned to each cluster defines the position of the cluster centroid. The samples are once again assigned to the new closest centroids and the process is repeated until convergence is reached. 

Fuzzy C-means is very similar to the K-means algorithm but 
the partition function $w_{i,j} \in [0,1]$ is given by
\begin{equation}
    w_{i,j} =\frac{1}{\sum_{j^\prime}^k \left( \frac{|| \vec{x}_i - \vec{c}_j ||}{|| \vec{x}_i - \vec{c}_{j^\prime} ||} \right)^\frac{2}{m-1}},
\end{equation}
where $m \in (0,\infty)$ is a parameter that controls the ``fuzziness'' of the clusters. 
These weights are a measure of the degree of membership of the sample $i$ to the cluster $j$ and now appear in the partition function, Eq.~(\ref{eq:partition}), to the power of $m$, as $(w_{i,j})^m$. Note that, contrary to K-means clustering, each sample can now be partially a member of several clusters simultaneously, to a certain degree given by the weight $w_{i,j}$. In the limit $m\rightarrow 1$, the membership converges to either $0$ or $1$ and we recover the K-means algorithm. The bigger the value of the fuzziness parameter $m$, the larger the overlap between clusters and, as a matter of fact, in the limit $m\rightarrow \infty$, all samples belong to all clusters equally. A benefit of this fuzzy approach is that it allows to distinguish the points that are more representative of each cluster and those that are in-between.

\begin{figure}[t]
    \centering
        \hspace{-15pt}
        \includegraphics[width=0.45\textwidth, trim={0 100 0 0},clip]
        {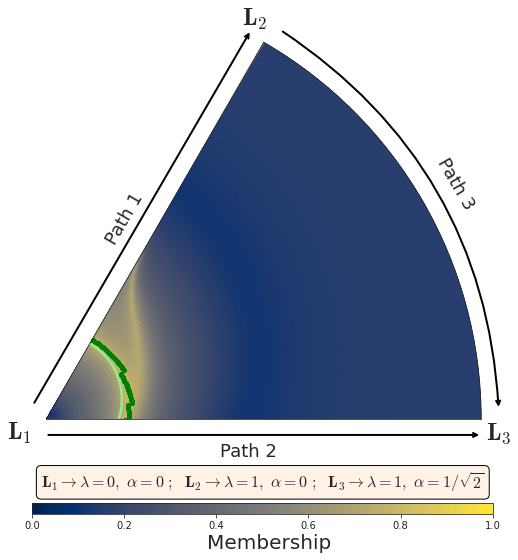} \\
        \hspace{-25pt}
        \includegraphics[width=0.56\textwidth, trim={0 0 0 460},clip]
        {Figures/ML/FuzzyCmeans/Fuzzy_Cmeans_maxes.png}
    \caption{Cluster 1 membership for the phase diagram of the ELM, computed through the Fuzzy C-means algorithm for a system size of $N=6$ (see text). 
    The light green line $\lambda_c = 1/(5+\alpha^2)$ corresponds to the theoretical critical points. The dark green line corresponds to the maximum value of the membership for each $\alpha$ value.}
    \label{fig:predictions_fuzzy}
\end{figure}

In Fig.~\ref{fig:predictions_fuzzy}, the value of the cluster membership for each point of the phase diagram is presented, using the Fuzzy C-means algorithm with $k=2$ clusters and a fuzziness parameter $m=4$. Since  $k=2$, $w_{i,1} = 1 - w_{i,2}$. Note that this algorithm is not capturing the quantum phases of the system, but whether a point is close to the critical area or the phase is well defined, without distinguishing between the symmetric or the broken phase. 
Thanks to this kind of soft clustering, we can identify the points that have a higher (or lower) degree of membership as the points closest to the QPT of the system. In fact, the dark green line in Fig.~\ref{fig:predictions_fuzzy} corresponds to the maximum value of the degree of membership for each value of $\alpha$, which is close to the theoretical critical points for each value of $\alpha$ (light green line). Notably, for a system with a single QPT and two phases, this prediction of the QPT line allows for a relatively precise reconstruction of the quantum phase diagram, even if the unsupervised ML algorithm is unable to explicitly distinguish the corresponding quantum phases.

In general, predicting the location of the QPT with unsupervised ML methods can become a difficult task, since the system is expected to behave very similarly when measured on two points if both of them are near the critical point, even though they may correspond to different phases. This is specially true for small system sizes, like the ones used on the supervised ML methods before, where due to finite size effects the QPT becomes smooth instead of sharp.

\section{Summary and conclusions}
\label{Sect-6}
In this work, we have explored the possibility of implementing the extended Lipkin model (ELM) in a analog-digital quantum simulator. First, we have presented the model and shown the connection with the nuclear interacting boson approximation (IBA). We have paid special attention to the connection between both models in the large-$N$ limit where the phase diagram is identical in the two cases. In both models, spherical and deformed phases exist and, depending on the trajectory followed in the phase space, first- and second-order quantum phase transitions appear. Furthermore, we have used the ADAPT-VQE algorithm to calculate the ground state energy. To this end, the Jordan-Wigner mapping is not needed, because the ELM in the spin representation is directly mapped into the quantum simulator. The results are very close to the exact ones, but the convergence strongly depends on the position of the Hamiltonian in the parameter space. Next, a proposal for the implementation of the time evolution operator of the ELM on a quantum platform based on the Digital-Analog Quantum Computation (DAQC) paradigm is presented. Calculations of the expected fidelities are also presented.
Finally, we present a Machine Learning study to identify the phases of the ELM. This study proposes, as proxies to identify the phase of the system, the time evolution of $\langle S_z (t) \rangle$ and a two-point correlator $C_z(1,2,t)$, Eq.\ (\ref{eq:correlation}).  Our analysis has three different flavours. We first perform a supervised classification task with full data. Then, we carry out the same exercise with partial, noisy information. Lastly, we also use an unsupervised classification exercise based on the C-means algorithm. In all cases, the results obtained capture the position of the different phases and can predict rather precisely the critical area. 

These studies lay the groundwork for the implementation of more intricate nuclear structure models on quantum platforms. In particular, the use of the ADAPT-VQE method to calculate the ground state energy or even the excited states opens the possibility of dealing with nuclear systems with large Hilbert spaces \cite{Pere2023}.

The proposed supervised method for determining the phase diagram of the system (see Sec.~\ref{Sect-5}) requires the calculation of the evolution of a well chosen observable for an arbitrary wave function that does not correspond to the ground state of the system. This is a very relevant fact, because the calculation of the ground state in medium and large size many-body systems can be very challenging. The method nonetheless provides accurate results using very simple states and taking advantage of Machine Learning techniques. Moreover, the use of unsupervised methods (see Sect.~\ref{sec-unsuper}) has provided very promising results allowing to locate the position of the critical area in the phase diagram with minimal biases.

\section{Acknowledgments}
This work was partially supported 
through the projects PID2022-136228NB-C21, PID2022-136228NB-C22, PID2020-114687GB-I00, PID2020-118758GB-I00 funded by MCIN/AEI/10.13039/50110001103 and by ``ERDF A way of making Europe''; 
``Ram\'on y Cajal" grant RYC2018-026072 funded by MCIN/AEI/10.13039/5011\-00011033 and by  “ESF Investing in your future”; the ``Consolidaci\'on Investigadora'' Grant CNS2022-135529 
funded by MCIN/AEI/10.13039/\-501100011033 and by the “European Union NextGenerationEU/PRTR” and the ``Unit of Excellence Mar\'ia de Maeztu 2020-2023" award to the Institute of Cosmos Sciences, Grant CEX2019-000918-M funded by MCIN/AEI/10.13039/501\-100011033.
This work has also been financially supported by the Ministry for Digital Transformation and of Civil Service of the Spanish Government through the QUANTUM ENIA project call - Quantum Spain project, and by the “European Union NextGenerationEU/PRTR” within the framework of the ``Digital Spain 2026 Agenda''. Resources supporting this work were provided by the CEAFMC and the Universidad de Huelva High Performance Computer (HPC@UHU) funded by ERDF/MINECO project UNHU-15CE-2848. 

\bibliography{references-other,references-IBM-CM,references-QPT,references-QC}
\end{document}